 \documentclass{jpp}

\usepackage{graphicx}
\usepackage{natbib}

\usepackage{epstopdf}
\usepackage{hyperref}

\usepackage{graphics}
\usepackage{epsfig}
\usepackage{bm}
\usepackage{amsmath,amssymb}
\usepackage{stmaryrd}

\usepackage{relsize}

\usepackage{arydshln}
\usepackage{mathtools}
\usepackage{amsfonts}
\usepackage{amsmath}
\usepackage{amssymb}
\usepackage{graphicx}
\usepackage{bm}
\usepackage{dcolumn}
\usepackage{color}
\usepackage{comment} 
\usepackage{amstext,graphics,subfigure}

\usepackage{txfonts}

\def\cala{\mathcal{A}}
\def\calb{\mathcal{B}}
\def\calc{\mathcal{C}}
\def\cald{\mathcal{D}}

\def\calh{\mathcal{H}}
\def\cali{\mathcal{I}}
\def\calj{\mathcal{J}}

\def\call{\mathcal{L}}

\def\calp{\mathcal{P}}
\def\calq{\mathcal{Q}}

\def\calt{\mathcal{T}}

\def\D{\mathbb{D}}
\def\Q{\mathbb{Q}}

\def\E{\mathbb{E}}

\def\P{\mathbb{P}}

\def\bq{\begin{equation}}
\def\eq{\end{equation}}
\def\bqy{\begin{eqnarray}}
\def\eqy{\end{eqnarray}}

\def\bal#1\eal{\begin{align}#1\end{align}}

\def\al{\alpha}
\def\be{\beta}

\def\de{\delta}
\def\De{\Delta}
\def\ep{\epsilon}

\def\ga{\gamma}
\def\Ga{\Gamma}

\def\la{\lambda}

\def\na{\nabla}

\def\rh{\rho}
\def\si{\sigma}

\def\bfB{\mathbf{B}}
\def\bfC{\mathbf{C}}
\def\bfD{\mathbf{D}}

\def\bfG{\mathbf{G}}
\def\bfA{\mathbf{A}}
\def\bfa{\mathbf{a}}

\def\bfJ{\mathbf{J}}
\def\bfL{\mathbf{L}}
\def\bfP{\mathbf{P}}

\def\bfM{\mathbf{M}}

\def\bfV{\mathbf{V}}
\def\bfn{\mathbf{n}}
\def\bfx{\mathbf{x}}

\def\bfv{\mathbf{v}}
\def\bfw{\mathbf{w}}
\def\bfW{\mathbf{W}}
\def\bfp{\mathbf{p}}
\def\bfq{\mathbf{q}}

\def\bfz{\mathbf{z}}

\def\bfGa{\boldsymbol{\Ga}}
\def\bfpi{\boldsymbol{\pi}}

\def\p{\partial}

\def\na{\nabla}

  \shorttitle{Lagrangian and Dirac constraints for fluid and MHD}
  \shortauthor{P. J. Morrison,   T. Andreussi,  and F. Pegoraro}

\title{Lagrangian and Dirac constraints for the ideal incompressible fluid and magnetohydrodynamics}

\author{P. J. Morrison\aff{1}
\corresp{\email{morrison@physics.utexas.edu}},
 T. Andreussi,\aff{2}    \and F. Pegoraro\aff{3}}
\affiliation{\aff{1}Department of Physics and  Institute for Fusion Studies, The University of Texas at Austin, Austin, TX 78712-1060, USA
\aff{2}SITAEL S.p.A., Pisa, 56121, Italy
\aff{3}Dipartimento di Fisica E.~Fermi, Pisa, 56127, Italy}

\begin{document}

\maketitle

\begin{abstract}
The incompressibility constraint for fluid flow was imposed by Lagrange in the so-called Lagrangian variable description  using his method of multipliers in the Lagrangian (variational) formulation.   An alternative is the imposition of   incompressibility in the Eulerian variable description by a generalization of Dirac's constraint method using noncanonical Poisson brackets. Here it  is shown how to impose the incompressibility constraint using Dirac's  method  in terms of both the canonical Poisson brackets in the Lagrangian variable description and the noncanonical Poisson brackets in the Eulerian description, allowing for the advection of density.  Both cases give dynamics of infinite-dimensional geodesic flow on the group of volume preserving diffeomorphisms and explicit expressions for this dynamics in terms of the constraints and original variables is given.  Because Lagrangian and Eulerian conservation laws are not identical,  comparison of the various methods is made.

\end{abstract}



\section{Introduction}
\label{sec:intro}

\subsection{Background}
\label{ssec:bgnd}

Sometimes constraints are maintained effortlessly, an example being $\nabla\cdot\bfB=0$  in electrodynamics which if initially true remains true,  while alternatively in most cases dynamical equations must be modified to maintain constraints, an example being $\nabla\cdot\bfv=0$ in fluid mechanics.  The need to apply constraints arises in a variety of contexts, ranging from gauge field theories \citep[e.g.][]{sundermeyer} to optimization and control \citep[e.g.][]{bloch}.  A very common approach is to use the method of Lagrange multipliers, which is taught in standard physics curricula for imposing holonomic constraints in mechanical systems. Alternatively,   \citet{dirac50}, in pursuit of his  goal of quantizing gauge field theories,  introduced a method that uses  the Poisson bracket.

The purpose of the present article is to explore different methods for imposing the compressibility constraint in ideal (dissipation-free)  fluid mechanics and its extension to magnetohydrodynamics (MHD),  classical field theories intended for classical purposes.  This endeavor is richer than might be expected because the different methods of constraint can be applied to the fluid  in either the Lagrangian  (material) description or the Eulerian (spatial) description, and the constraint methods  have different manifestations in the Lagrangian (action principle) and Hamiltonian formulations.    Although Lagrange's multiplier is widely appreciated, it is not so well known that he used it  long ago for imposing the incompressibility constraint for a fluid in the Lagrangian variable description  \citep{lagrange}.  More recently, Dirac's method was applied for imposing incompressibility within the Eulerian variable description,  first in  \citet{turski99,turski01} and followed up in several works \citep{pjmLB09,pjmTC09,pjmCT12,pjmCT14,pjmCGBT13}.   Given that a Lagrangian conservation law is not equivalent to an Eulerian conservation law, it remains to elucidate the interplay between the methods of constraint and the variables used for the description of the fluid.  Thus we have three dichotomies: the Lagrangian vs.\ Eulerian fluid descriptions, Lagrange multiplier vs.\ Dirac constraint methods, and Lagrangian vs.\ Hamiltonian formalisms. It is the elucidation of the interplay between these, along with generalizing previous results, that is the present goal. 

It is well known that a free particle with  holonomic constraints,  imposed by the method of Lagrange multipliers,  is a geodesic flow.  Indeed, Lagrange essentially observed this in  \citet{lagrange} for the ideal fluid when he imposed the incompressibility constraint  by his method. Lagrange did this in the Lagrangian description by imposing the constraint that the map from initial positions of fluid elements to their positions at time $t$ preserves  volume, and he did this by the method of Lagrange multipliers.  It is worth noting that  Lagrange  knew the  Lagrange multiplier turns out to be the pressure, but he had trouble solving for it.   Lagrange's calculation was placed in a geometrical setting by \citet{arnold-diffeo} (see also Appendix 2 of \cite{arnold-book}), where the constrained maps from the initial conditions were first  referred to as volume preserving diffeomorphisms in this context.  Given this background, in our investigation of the three dichotomies described above we emphasize geodesic flow.   
 
For later use we record here the incompressible Euler equations for the case with constant density and the case where  density is advected.  The equations of motion, allowing for density advection,   are given by
\bal
\frac{\p \bfv}{\p t} &= -\bfv\cdot\na\bfv -\frac1{\rho}\na p\,,
\label{momentum}
\\
& \na \cdot \bfv=0\,,
\label{solenoidal}
\\
\frac{\p \rho}{\p t} &= -\bfv\cdot\na\rho\,,
\eal
where $\bfv(\bfx,t)$ is the velocity field, $\rho(\bfx,t)$ is the mass density,  $p(\bfx,t)$ is the pressure, and $\bfx\in D$, the region occupied by the fluid. These equations are generally subject to the  free-slip boundary condition $\bfn\cdot \bfv|_{\p D}=0$, where $\bfn$ is  normal to the boundary of $D$.    The pressure field that enforces the  constraint  \eqref{solenoidal} is obtained by setting $\p (\nabla\cdot \bfv)\p t=0$, 
which implies
\bq
\De_\rh p:=\nabla\cdot\left(\frac1{\rho}\nabla p\right)=-\na \cdot (\bfv\cdot\na\bf v)\,.
\label{elliptic}
\eq
For reasonable assumptions on $\rho$ and boundary conditions, \eqref{elliptic} is a well-posed elliptic equation \citep[see e.g.][] {evans}, so we can write
\bq
p= -\De_\rh^{-1}\nabla\cdot(\bfv\cdot\na\bf v)\,.
\label{presrho}
\eq
For the case where  $\rho$ is constant we have the usual Green's function expression
\bq
p(\bfx,t)=- \, \rho \int \! d^3x' \, G(\bfx,\bfx')\, \na'\! \cdot (\bfv'\cdot\na'\bf v')\,,
\label{pG}
\eq
where $G$ is consistent with  Neumann boundary conditions \citep{orszag86} and $\bfv'=\bfv(\bfx',t)$.    Insertion of \eqref{pG} into \eqref{momentum} gives
\bq
\frac{\p \bfv}{\p t} = -\bfv\cdot\na\bfv +\na\int\! d^3x' \, G(\bfx,\bfx')\,  \na'\!\cdot (\bfv'\cdot\na'\bf v')\,, 
\label{momentumCl}
\eq
which is a closed system for $\bfv(\bfx,t)$. 

For MHD,    equation \eqref{momentum} has the additional term $(\nabla\times\bfB)\times \bfB/\rho$ added to the righthand side.  Consequently for this model, the source of \eqref{presrho} is modified by the addition of this term to $-\bfv\cdot\na\bf v$. 
\subsection{Overview}
\label{ssec:over}

Section \ref{sec:constraints} contains material that serves as a guide for navigating the more complicated calculations to follow.   We first consider the various approaches to constraints in the finite-dimensional context in  Sections \ref{ssec:lagrange}--\ref{ssec:holoD}.  Section  \ref{ssec:lagrange} briefly covers  conventional material about holonomic constraints by Lagrange multipliers --  here the reader is reminded  how  the free particle with holonomic constraints  amounts to geodesic flow.  Section \ref{ssec:dirac} begins with the phase space action principle, whence the Dirac bracket for constraints  is obtained by Lagrange's multiplier method, but with phase space constraints as opposed to the usual holonomic configuration space constraints used in conjunction with  Hamilton's principle of mechanics,  as described in Section  \ref{ssec:lagrange}.  Next,  in Section \ref{ssec:holoD}, we compare the results of Sections \ref{ssec:lagrange} and  \ref{ssec:holoD} and show how conventional holonomic constraints can be enforced by Dirac's method.   Contrary to Lagrange's method,  here we obtain explicit expressions, ones that do not appear in conventional treatments,  for the Christoffel symbol and the normal force entirely in terms of the original Euclidean coordinates and constraints. Section  \ref{sec:constraints}  is completed  with Section \ref{ssec:d+1}, where the previous ideas are revisited in the $d+1$  field theory context in  {preparation} for the fluid and MHD calculations.  Holonomic constraints,  Dirac brackets, with local or nonlocal constraints, and geodesic flow are treated.

In Section \ref{sec:unconstrained} we first consider the compressible (unconstrained) fluid and MHD versions of Hamilton's variational principle, the principle of least action,  with Lagrange's Lagrangian in the Lagrangian description.  From this we obtain in Section \ref{ssec:HamDesc} the canonical Hamiltonian field theoretic form in the Lagrangian variable description, which is transformed in Section \ref{ssec:LtoE}, via the mapping from Lagrangian to Eulerian variables, to the noncanonical Eulerian form.  Section  \ref{sec:unconstrained} is completed by an in depth  comparison of constants of motion in the Eulerian and Lagrangian descriptions, which surprisingly does not seem to appear in fluid mechanics or plasma physics textbooks.  The material of this section is necessary for understanding the different manifestations of constraints in our dichotomies. 

Section \ref{sec:dirac} begins with Section \ref{ssec:LagCon} that reviews Lagrange's original calculations. Because the incompressibility constraint he imposes is holonomic and there are no additional forces, his equations  describe infinite-dimensional geodesics flow on volume preserving maps.  The remaining portion of this section contains the most substantial calculations of the paper.  In Section  \ref{ssec:LDconTh} for the first time Dirac's theory is applied to enforce incompressibility in the Lagrangian variable description.  This results in a new Dirac bracket that generates volume preserving flows. As in Section \ref{sssec:HCD}, which serves as a guide, the  equations of motion generated by the bracket  are explicit and contain only the constraints and original variables.  Next, in Section \ref{ssec:EDcon}, a  reduction from Lagrangian to Eulerian variables is made, resulting in a new Eulerian variable Poisson bracket that allows for density advection while preserving incompressibility.  This was an heretofore unsolved problem.  Section \ref{ssec:comparison} ties together the results of Sections  \ref{ssec:LDconTh}, \ref{ssec:EDcon}, and \ref{ssec:CoM}. Here both the Eulerian and Lagrangian Dirac constraint theories are compared after they are evaluated on their respective constraints, simplifying their equations of motion.  Because Lagrangian and Eulerian conservation laws are not identical, we see that there are differences.   Section  \ref{sec:dirac} concludes in Section \ref{ssec:AoI} with a discussion of the full algebra of invariants, that of the ten parameter Galilean group, for both the Lagrangian and Eulerian descriptions.  In addition the  Casimir invariants of the theories are discussed. 

The paper concludes with  Section \ref{sec:conclusion}, where we briefly summarize our results and speculate about future possibilities.

\section{Constraint methods}
\label{sec:constraints}

\subsection{Holonomic constraints by Lagrange's multiplier method}
\label{ssec:lagrange}

 Of interest are systems with Lagrangians of the form $L(\dot{q}, q)$ where the overdot denotes time differentiation and $q=(q^1,q^2,\dots, q^N)$.  Because nonautonomous systems could be included by appending an additional degree of freedom,  explicit time dependence is not included in $L$. 
 
 Given the Lagrangian, the equations of motion are obtained according to Hamilton's principle by variation of the action
 \bq
 S[q]=\int_{t_0}^{t_1}\!\!dt\, L(\dot{q}, q)\,;
 \label{HamPrin}
 \eq
i.e.
\bq
\de S[q;\de q]:= \frac{d}{d\ep} S[q +\ep \de q]\Big|_{\ep=0} 
=\int_{t_0}^{t_1}\!\!dt\left(  \frac{d }{dt} \frac{\p L}{\p \dot{q}^i} - \frac{\p L}{\p {q}^i}\right)\de q^i
=\int_{t_0}^{t_1}\!\!dt\, \frac{\de S[q]}{\de q^i(t)}\, \de q^i= 0\,, 
\eq
for all variations $\de q(t)$ satisfying $\de q(t_0)=\de q(t_1)=0$, implies  Lagrange's equations of motion, i.e. 
 \bq
 \frac{\de S[q]}{\de q^{i}(t)} = 0\quad \Rightarrow\quad \frac{d }{dt} \frac{\p L}{\p \dot{q}^i} - \frac{\p L}{\p {q}^i}=0\,, \qquad i=1,2,\dots, N\,.
 \label{LagEOM}
 \eq
 
Holonomic constraints are real-valued functions of the form $C^A(q)$,  $A=1,2,\dots, M$,  which are desired to be constant on trajectories.   Lagrange's method for implementing such constraints is to add them to the action and vary as follows:
 \bq
  \de S_\la :=\de (S + \la_A C^A)=0\,, 
 \label{Lagconstraint}
 \eq
yielding the  equations of motion
 \bq
  \frac{\de S_{\la}[q]}{\de q^{i}(t)} = 0\quad \Rightarrow\quad
  \frac{d }{dt} \frac{\p L}{\p \dot{q}^i} - \frac{\p L}{\p {q}^i}=\la_A\frac{\p C^A}{\p q^i}\,, \qquad i=1,2,\dots, N\,, 
 \label{LagForce}
 \eq
with the forces of constraint residing on the right-hand side of \eqref{LagForce}.  Observe in  \eqref{Lagconstraint} and \eqref{LagForce} repeated sum notation is implied for the index $A$. The $N$ equations of \eqref{LagForce} with the $M$ numerical values of the constraints $C^A(q)=C_0^A$, determine the  $N+M$ unknowns $\{q^i\}$ and $\{\la_A\}$.  In practice, because solving for the Lagrange multipliers can be difficult an alternative procedure, a example of which we describe in Section \ref{sssec:HG},  is used. 
 
   We will see in Section \ref{ssec:LagCon} that the field theoretic version of this method is how Lagrange implemented the incompressiblity constraint for fluid flow.  For the purpose of illustration and in preparation for later development, we consider a finite-dimensional analog of Lagrange's treatment.

\subsubsection{Holonomic constraints and geodesic flow via Lagrange}
\label{sssec:HG}

Consider  $N$ noninteracting bodies each of mass $m_i$  in the   Eucledian configuration space $\E^{3N}$ with cartesian coordinates $\bfq_i=(q_{xi},q_{yi},q_{zi})$,  {where as in  Section  \ref{ssec:lagrange}  $i=1,2,\dots,N$, but our configuration space has dimension $3N$.} The Lagrangian for this system is given by the  usual kinetic energy,
   \bq
   L= \sum_{i=1}^N \frac{m_i}{2} \, \dot{\bfq}_i\cdot  \dot{\bfq}_i\,,
   \label{fpl}
   \eq
with the usual ``dot" product.  The Euler-Lagrange equations for this system, equations \eqref{LagEOM},  are  the uninteresting system of $N$ free particles.   As in Section \ref{ssec:lagrange} we constrain this system by adding constraints  $C^A(\bfq_1,\bfq_2,\dots,\bfq_N)$,  where again $A=1,2,\dots, M$, leading to  the equations
\bq
m_i\ddot{\bfq}_i = \la_A \frac{\p C^A}{\p \bfq_i}\,.
\label{ELgeod}
\eq

Instead of solving the $3N$ equations of \eqref{ELgeod} together with the $M$ numerical values of the constraints, in order to determine the  unknowns $\bfq_i$ and $\la_A$,  we recall the alternative procedure, which dates back to Lagrange (see e.g.\ Sec. IV of \citet{lagrange}) and has been taught to physics students for generations \citep[see e.g.][] {whittaker,corben}.  With the alternative procedure one introduces generalized coordinates that account for the constraints, yielding a smaller system on the constraint manifold, one with the Lagrangian
\bq
L=\frac12 g_{\mu\nu}(q) \, \dot{q}^\mu \dot{q}^\nu\,, \hspace{1cm} \mu,\nu= 1,2\dots, 3N-M\,,
\label{laggeo}
\eq
where 
\bq
g_{\mu\nu}= \sum_{i=1}^Nm_i \frac{\p \bfq_i}{\p q^{\mu}}\cdot\frac{\p \bfq_i}{\p q^{\nu}}
\,.
\label{gdef}
\eq
Then  Lagrange's equations \eqref{LagEOM} for the Lagrangian \eqref{laggeo} are the usual equations for geodesic flow
\bq
\ddot{q}^\mu+\Ga^\mu_{\al\be}\, \dot{q}^\al\dot{q}^\be=0\,,
\label{geoflow}
\eq
where the Christoffel symbol is as usual 
\bq
  \Ga^\mu_{\al\be}=\frac12 g^{\mu\nu}\left(
 \frac{g_{\nu\al}}{\p q^\be} +  \frac{g_{\nu\be}}{\p q^\al} -  \frac{g_{\al\be}}{\p q^\nu} 
 \right)
\,.
\label{Csymb}
\eq
 If the constraints had time dependence, then the procedure would have produced  the Coriolis and centripetal forces, as is usually done in textbooks. 
 
 Thus, we arrive at the conclusion that free particle dynamics with time-independent holonomic constraints is geodesic flow.

\subsection{Dirac's bracket method}
\label{ssec:dirac}

So, a natural question to ask is ``How  does one implement constraints in the Hamiltonian setting, where phase space constraints  depend on both the configuration space coordinate $q$ and the  canonical momentum $p$"?  {(See e.g.\ \cite{sundermeyer,AKN} for a general treatment and \cite{moncrief}  for a treatment  in  the context of the ideal fluid and  relativity and a selection of earlier references.) } To this end we begin with  the phase space action principle
\begin{equation}
S[q,p]=\int_{t_0}^{t_1}\!\!dt\, \left[  p_i    \dot{q}^i  - H(q,p)\right]\,,
\end{equation}
where again repeated sum notation is used for $i=1,2,\dots, N$.  Independent  variation of $S[q,p]$   with   respect to $q$ and $p$, with $\de q(t_0)=\de q(t_1)=0$ and no conditions on $\de p$,  yields Hamilton's equations,  
\bq
\dot{p}_i =-\frac{\partial H}{\partial q^i} \qquad \mathrm{and}\qquad
\dot{q}^i  = \frac{\partial H}{\partial p_i}  \,, 
\label{finite-ham}
\eq
or equivalently
\bq
 \dot{z}^{\al}=\{z^\al,H\} \,,
 \eq
which is a rewrite of \eqref{finite-ham} in terms of the Poisson bracket on phase space functions $f$ and $g$, 
 \bq
 \{ f,g\} =\frac{\p f}{\p q^i}\frac{\p g}{\p p_i}- \frac{\p g}{\p q^i}\frac{\p f}{\p p_i} 
 = \frac{\p f}{\p z^\al}\mathbb{J}_c^{\al\be}\frac{\p g}{\p z^\be}\,,
 \label{FDcanbkt}
 \eq
where in the second equality we have used  $z=(q,p)$,  so     $\al,\be=1,2,\dots, 2N$ and the cosymplectic form (Poisson matrix) is
\bq
\mathbb{J}_c  = \left( \begin{array}{cc}
\mathbb{O}_N  &  \mathbb{I}_N  
\\
-\mathbb{I}_N  &  \mathbb{O}_N   \end{array} \right) \,, 
\label{Jc}
\end{equation}
 with $\mathbb{O}_N$ being an $N\times N$ block of zeros and $\mathbb{I}_N$ being the $N\times N$ identity. 
 
Proceeding as in Section \ref{ssec:lagrange},  albeit with phase space constraints $D^a(q,p)$, $a=1,2,\dots, 2M< 2N$, we vary 
\begin{equation}
S_\la[q,p]=\int_{t_0}^{t_1}\!\!dt\, \left[  p_i \dot{q}^i -H(q,p) + \lambda_a 
D^a\right]\,,
\end{equation}
and  obtain 
\bq
\dot{p}_i =- \frac{\partial H}{\partial q^i}  
+  \lambda_a \frac{\partial D^a}{\partial q^i} 
\qquad \mathrm{and}\qquad 
\dot{q}^i =\frac{\partial H}{\partial p_i}  
- \lambda_a \frac{\partial D^a}{\partial p_i}\,.
\label{condyn}
\eq
Next, enforcing  $\dot D^a=0$ for all $a$, will ensure that the constraints  stay put.  Whence,   differentiating the $D^a$ and using \eqref{condyn}   yields 
 \begin{eqnarray}
 \dot{D}^a&=&   \frac{\partial D^a}{\partial q^i}{\dot q}^i + 
  \frac{\partial D^a}{\partial p_i}{\dot p}_i \nonumber\\
&=&\{D^a,H\}  -    \la_b \{D^a,D^b\} \equiv 
0\,.
   \label{Cdyn} 
 \end{eqnarray}
   We assume  $\D^{ab}:=\{D^a,D^b\}$ has an inverse, $\D^{-1}_{ab}$,  which requires there be an even number of constraints, $a,b=1,2,\dots, 2M$,  because odd antisymmetric matrices have zero determinant. Then upon solving 
(\ref{Cdyn}) for $\lambda_b$ and inserting the result into \eqref{condyn}   gives
 \begin{equation}
 \dot{z}^{\al}=\{z^\al,H\} - \D^{-1}_{ab}\{z^\al,D^{a}\}\{D^b,H\}\,.
 \label{czdyn}
 \end{equation}
From (\ref{czdyn}), we obtain a generalization of the Poisson bracket,  the Dirac bracket, 
\begin{equation}
\{f,g\}_*=\{f,g\} - \D^{-1}_{ab}\{f,D^{a}\}\{D^b,g\}\,.
\label{DB}
\end{equation}
which has the degeneracy property 
\bq
\{f,D^a\}_*\equiv 0\  \,.
\label{DCas}
\eq
for all functions $f$ and indices $a=1,2,\dots, 2M$.

The generation of the equations of motion via a Dirac bracket, i.e.
\bq
\dot{z}^{\al}=\{z^\al,H\}_*\,, 
\label{evolv}
\eq
which is equivalent to \eqref{czdyn}, has the advantage that the Lagrange multipliers $\la_A$ have been eliminated from the theory. 

Note, although the above construction of the Dirac bracket is based on the canonical bracket of \eqref{FDcanbkt}, his construction results in a valid Poisson bracket if one starts from any valid Poisson bracket (cf.\  \eqref{eqn:PBgene} of Section \ref{ssec:d+1} and Section  \ref{ssec:LtoE}), which need not have a Poisson matrix of the form of \eqref{Jc} \citep[see e.g.] []{pjmLB09}.  We will use such a bracket in Section \ref{ssec:EDcon} when we apply constraints by Dirac's method in the Eulerian variable picture.  Also note, for our purposes  it is not necessary to  describe  primary vs.\  secondary constraints (although we use the latter), and the notions of weak vs.\  strong equality.  We refer the reader to  \citet{dirac50,sudarshan,HRT,sundermeyer} for treatment of these concepts.

 \subsection{Holonomic constraints by Dirac's bracket method}
\label{ssec:holoD}

 A connection between the approaches of Lagrange and Dirac can be made.   From a set of  Lagrangian constraints $C^A(q)$, where $A=1,2,\dots,M$, one can construct an additional $M$ constraints by differentiation, 
 \bq
 \dot{C}^A=\frac{\p C^A}{\p q^i} \dot{q}^i= \frac{\p C^A}{\p q^i}\frac{\p H}{\p p_i}\,, 
 \eq
 where the second equality is possible if \eqref{LagEOM}   possesses the Legendre transformation to the Hamiltonian form.  In this way we obtain an even number of constraints
 \bq
 D^a(q,p)= \left(C^A(q)\,, \dot{C}^{A'}(q,p)\right)\,,
\label{LDcon}
\eq 
where $A= 1,2,\dots, M$, $A'= M+1, M+2,\dots, 2M$, and $a=1,2,\dots, 2M.$  

 With the constraints of \eqref{LDcon} the  bracket $\D^{ab}= \{D^a,D^b\}$ needed to construct  \eqref{DB} is easily obtained, 
 \bq
\D
  = \left( \begin{array}{cc}
\mathbb{O}_M  &  \{C^A, \dot{C}^{B'}\}  
\\
 \{\dot{C}^{A'},C^{B}\}    &   \{\dot{C}^{A'},\dot{C}^{B'}\}   \end{array} \right) =: 
 \left( \begin{array}{cc}
\,\mathbb{O}_M  &  \mathbb{S} 
\\
\!-\mathbb{S}   &  \mathbb{A}   \end{array} \right)\,,
\label{bbD}
\end{equation}
 where $\mathbb{O}_M$ is an $M\times M$ block of zeros and $ \mathbb{S}$ is the following $M\times M$ symmetric matrix with elements 
 \bq
\mathbb{S}^{AB}:=   \{C^A, \dot{C}^{B}\}  =
 \frac{\p^2 H}{\p p_i \p p_j}\,  \frac{\p C^A}{\p q^i}  \frac{\p C^B}{\p q^j}
 \,,
  \label{DAB}
 \eq
 and $\mathbb{A}$ is the following $M\times M$ antisymmetric matrix with elements
 \bal
\mathbb{A}^{AB}&:= \{\dot{C}^{A'},\dot{C}^{B'}\} 
 \label{D2D2}\\
&=\frac{\p^2 H}{\p p_i \p p_k}\left[
\frac{\p^2 H}{\p q^i \p p_j} 
 \left(
 \frac{\p C^A}{\p q^j} \frac{\p C^B}{\p q^k} -  \frac{\p C^B}{\p q^j} \frac{\p C^A}{\p q^k}
 \right)
 +
 \frac{\p H}{\p p_j}
 \left(
 \frac{\p^2 C^A}{\p q^i \p q^j} \frac{\p C^B}{\p q^k} -   \frac{\p^2 C^B}{\p q^i \p q^j} \frac{\p C^A}{\p q^k} 
 \right)
 \right]\,.
 \nonumber
 \eal

 Assuming the existence of $\D^{-1}$, the $2M\times 2M $ inverse of \eqref{bbD},   the Dirac bracket of \eqref{DB} can be constructed.   A necessary and sufficient condition for the existence of this inverse is that  det\,$\mathbb{S}\neq 0$, and when this is the case the inverse is given by 
 \bq
\D^{-1}
 =
 \left( \begin{array}{cc}
\mathbb{S}^{-1}\! \mathbb{A}\mathbb{S}^{-1}  & - \,\mathbb{S}^{-1} 
\\
\\
\, \mathbb{S}^{-1}   &  \mathbb{O}_M   \end{array} \right)\,.
\label{bbDI}
\eq
 Because of the block diagonal structure of \eqref{bbDI}, the Dirac bracket \eqref{DB} becomes
\bq
\{f,g\}_*=\{f,g\} + \mathbb{S}^{-1}_{AB}\left(
\{f,C^{A}\}\{\dot{C}^{B},g\} -\{g,C^{A}\}\{\dot{C}^{B},f\} 
\right) + \mathbb{S}^{-1}_{AC}\, \mathbb{A}^{CD}\, \mathbb{S}^{-1}_{DB}\, \{f,{C}^{A}\}\{{C}^{B},g\}\,, 
\label{DBbb}
\eq
which has the form
\bal
\{f,g\}_*&=\{f,g\} -  (\P_\perp)^i_j\left( \frac{\p f}{\p q^i}\frac{\p g}{\p p_j} - \frac{\p g}{\p q^i}\frac{\p f}{\p p_j}\right) 
+ \Q^{ij} \,  \frac{\p f}{\p p_i}\frac{\p g}{\p p_j} 
\nonumber\\
&= \frac{\p f}{\p q^i} \, \P^i_j\frac{\p g}{\p p_j} -  \frac{\p g}{\p q^i}\,  \P^i_j\frac{\p f}{\p p_j}
+ \Q^{ij}\,  \frac{\p f}{\p p_i}\frac{\p g}{\p p_j} 
\,, 
\label{DBbbP}
\eal
where the matrices $\P=I-\P_{\perp}$, with 
\bq
(\P_{\perp})_j^i =  \mathbb{S}^{-1}_{AB} \, \frac{\p^2 H}{\p p_i \p p_k} 
 \frac{\p C^A}{\p q^j} \frac{\p C^B}{\p q^k} \,,
\eq
 and $\Q$, a complicated expression that we will not record,  are crafted using the constraints and Hamiltonian so as to make $\{f,g\}_*$ preserve the constraints. 

The equations of motion that follow from \eqref{DBbb} are 
\bal
\dot{q}^{\ell}&=\{q^{\ell},H\}_*=\frac{\p H}{\p p_\ell}+ \mathbb{S}^{-1}_{AB}\left(
\{q^{\ell},C^{A}\}\{\dot{C}^{B},H\} -\{H,C^{A}\}\{\dot{C}^{B},q^{\ell}\} 
\right) 
\nonumber\\
&\hspace{3cm} + \mathbb{S}^{-1}_{AC}\, \mathbb{A}^{CD}\, \mathbb{S}^{-1}_{DB}\, \{q^{\ell},{C}^{A}\}\{{C}^{B},H\}\,,
\label{eomq}\\
\dot{p}_{\ell}&=\{p_{\ell},H\}_*=-\frac{\p H}{\p q^\ell}+ \mathbb{S}^{-1}_{AB}\left(
\{p_{\ell},C^{A}\}\{\dot{C}^{B},H\} -\{H,C^{A}\}\{\dot{C}^{B},p_{\ell}\} 
\right) 
\nonumber\\
&\hspace{3cm}  + \mathbb{S}^{-1}_{AC}\, \mathbb{A}^{CD}\,\mathbb{S}^{-1}_{DB}\, \{p_{\ell},{C}^{A}\}\{{C}^{B},H\}\,.
\label{eomp}
\eal

 Given the Dirac bracket associated with the $\D$ of \eqref{DAB},  dynamics that enforces the constraints takes the form of  \eqref{evolv}.  Any  system generated by this bracket will  enforce Lagrange's holonomic constraints; however,  only initial conditions compatible with  
 \bq
 D^a\equiv 0\,, \qquad \forall \qquad  a=M+1, M+2, \dots, 2M\,,
\eq
or equivalently
 \bq
\dot{C}^A= \frac{\p C^A}{\p q^i}\frac{\p H}{\p p_i}=\{C^A,H\}\equiv 0\,, \qquad \forall  \qquad  A=1, 2, \dots, M\,,
\label{dotcon}
\eq
 will correspond to the  system with holonomic constraints.   Using \eqref{dotcon} and $\{q^{\ell},C^{A}\}\equiv 0$,  \eqref{eomq} and \eqref{eomp} reduce to 
 \bal
\dot{q}^{\ell}&=\{q^{\ell},H\}_*=\frac{\p H}{\p p_\ell}\,,
\label{eomqr}\\
\dot{p}_{\ell}&=\{p_{\ell},H\}_*=-\frac{\p H}{\p q^\ell}+ \mathbb{S}^{-1}_{AB} 
\{p_{\ell},C^{A}\}\{\dot{C}^{B},H\}  \,,
\label{eompr}
\eal
where
 \bq
 \{\dot{C}^{B},H\}=\left(\frac{\p^2 H}{\p q^i\p p_j}\frac{\p C^B}{\p q^j} + \frac{\p H}{\p p_j}\frac{\p^2 C^B}{\p q^iq^j}\right) \frac{\p H}{\p p_i} 
- \frac{\p C^B}{\p q^i} \frac{\p^2 H}{\p p_i\p p_j} \frac{\p H}{\p q^j}\,.
\label{dotCH}
\eq
Thus the Dirac bracket approach gives a relatively simple system for enforcing holonomic constraints.  It can be shown directly that if initially $\dot{C}^A$ vanishes, then the  system of \eqref{eomqr} and \eqref{eompr} will keep it so for all time. 
 
\subsubsection{Holonomic constraints and geodesic flow via Dirac}
\label{sssec:HCD}
  
Let us now consider  again the geodesic flow problem of Section \ref{sssec:HG}:  the  $N$ degree-of-freedom free particle system with holonomic constraints, but this time within the framework of Dirac bracket theory.   For this problem the unconstrained configuration   space is the Euclidean space $\E^{3N}$ and we will denote  by $\calq$ the constraint manifold within $\E^{3N}$ defined by the constancy of the constraints  $C^A$.  

The Lagrangian of \eqref{fpl} is easily Legendre transformed to the free particle Hamiltonian
\bq
   H= \sum_{i=1}^N \frac1{2m_i} \,  {\bfp}_i\cdot   {\bfp}_i\,,
   \label{fph}
   \eq
where $\bfp_i=m_i\dot{\bfq}_i$.    For this example the constraints of \eqref{LDcon} take the form
 \bq
 D^a= \left(C^A(\bfq_1,\bfq_2,\dots,\bfq_N),
 \frac{\p C^{A'}(\bfq_1,\bfq_2,\dots,\bfq_N)}{\p \bfq_i}
  \cdot \frac{\bfp_i}{m_i}
  \right)\,, 
 \eq
 the $M\times M$ matrix $\mathbb{S}$ has elements 
 \bq
 \mathbb{S}^{AB} = \sum_{i=1}^N\, \frac1{m_i} 
  \frac{\p C^A}{\p \bfq_i} \cdot \frac{\p C^B}{\p \bfq_i}
 \,, 
  \label{fpDAB}
 \eq
and  the $M\times M$ matrix $\mathbb{A}$ is 
 \bq
 \mathbb{A}^{AB} = \sum_{i,j=1}^N\, \frac1{m_i m_j} \,\bfp_i \cdot
 \left[
 \frac{\p^2 C^A}{\p \bfq_i \p \bfq_j}\cdot \frac{\p C^B}{\p \bfq_j}   
 -  \frac{\p^2 C^B}{\p \bfq_i \p \bfq_j}\cdot \frac{\p C^A}{\p \bfq_j}
 \right]\,.
 \eq
 The Dirac bracket analogous to \eqref{DBbbP} is 
 \bq
\{f,g\}_*=  \sum^N_{ij=1}\,\left[   \frac{\p f}{\p \bfq_i} \cdot \overset\leftrightarrow{\mathbb{P}}_{ij}  \cdot \frac{\p g}{\p \bfp_j} 
-  \frac{\p g}{\p \bfq_i} \cdot  \overset\leftrightarrow{\mathbb{P}}_{ij} \cdot \frac{\p f}{\p \bfp_j}
+  \frac{\p f}{\p \bfp_i}\cdot \overset\leftrightarrow{\mathbb{Q}}_{ij} \cdot \frac{\p g}{\p \bfp_j} 
\right]\,,
\label{geoBkt}
\eq
 where 
 $\overset\leftrightarrow{\mathbb{P}}_{ij} 
 = \overset\leftrightarrow{\mathbb{I}}_{ij} - \overset\leftrightarrow{\mathbb{P}}_{\perp\,  ij}$ with the tensors 
 \bal
\overset\leftrightarrow{\mathbb{P}}_{\perp\,  ij}&:= 
 \sum^N_{k=1}\,\mathbb{S}^{-1}_{AB}\,  \frac{\p^2 H}{\p \bfp_i \p \bfp_k} \cdot \frac{\p C^B}{\p \bfq_k}
 \frac{\p C^A}{\p \bfq_j} =  \mathbb{S}^{-1}_{AB}\,\frac1{m_i} \frac{\p C^B}{\p \bfq_i}   \frac{\p C^A}{\p \bfq_j}\,,
 \\
 \overset\leftrightarrow{\mathbb{Q}}_{\,  ij}&:=  \sum^N_{k=1}\,\mathbb{S}^{-1}_{AB}\,
 \left[
  \frac{\p C^A}{\p \bfq_j}\,  \frac{\bfp_k}{m_k} \cdot \frac{\p^2 C^B}{\p \bfq_k \p \bfq_i}
  -   \frac{\p C^A}{\p \bfq_i} \, \frac{\bfp_k}{m_k} \cdot \frac{\p^2 C^B}{\p \bfq_k \p \bfq_j}
  \right] 
  +   \mathbb{S}^{-1}_{AC}\, \mathbb{A}^{CD}\, \mathbb{S}^{-1}_{DB}\, \frac{\p C^{A}}{\p \bfq_i}  \frac{\p C^B}{\p \bfq_j}  
  \\
&=: \overset\leftrightarrow{\mathbb{T}}_{ij} - \overset\leftrightarrow{\mathbb{T}}_{ji}   
+\overset\leftrightarrow{\mathbb{A}}_{ij} \,,
\label{TA}
 \eal
 where $\overset\leftrightarrow{\mathbb{A}}_{ij}$ is the term with 
 $\mathbb{S}^{-1}_{AC} \mathbb{A}^{CD} \mathbb{S}^{-1}_{DB}$.
Observe  $\overset\leftrightarrow{\mathbb{A}}_{ij}= -\overset\leftrightarrow{\mathbb{A}}_{ji}$ because  $\mathbb{A}^{CD}=- \mathbb{A}^{DC}$ and 
\bal
\sum_{k=1}^N\, \overset\leftrightarrow{\mathbb{P}}_{\perp\,  ik} \cdot \overset\leftrightarrow{\mathbb{P}}_{\perp\,  kj}
&= \sum_{k=1}^N\,\left(
\mathbb{S}^{-1}_{AB}\,\frac1{m_i} \frac{\p C^B}{\p \bfq_i}   \frac{\p C^A}{\p \bfq_k}
\right)
\cdot
\left(
\mathbb{S}^{-1}_{A'B'}\,
\frac1{m_k} \frac{\p C^{B'}}{\p \bfq_k}  \frac{\p C^{A'}}{\p \bfq_j}
\right)
\nonumber\\
&= \left(
\mathbb{S}^{-1}_{AB}\,\frac1{m_i} \frac{\p C^B}{\p \bfq_i}\right) \mathbb{S}^{-1}_{A'B'}\, \left(  \sum_{k=1}^N\,  \frac{\p C^A}{\p \bfq_k}
\cdot
\frac1{m_k} \frac{\p C^{B'}}{\p \bfq_k}\right)
  \frac{\p C^{A'}}{\p \bfq_j}
  \nonumber\\
  &=  \left(
\mathbb{S}^{-1}_{AB}\,\frac1{m_i} \frac{\p C^B}{\p \bfq_i}\right) \mathbb{S}^{-1}_{A'B'}\, \mathbb{S}^{A B'}
  \frac{\p C^{A'}}{\p \bfq_j} = \overset\leftrightarrow{\mathbb{P}}_{\perp\,  ij} \,.
\eal
Also observe for the Hamiltonian of \eqref{fph}
 \bal
 \sum^N_{j=1}\, \overset\leftrightarrow{\mathbb{P}}_{\perp\, ij}  \cdot \frac{\p H}{\p \bfp_j}&=
 \sum^N_{j=1}\, \overset\leftrightarrow{\mathbb{P}}_{\perp\, ij}  \cdot\frac{\bfp_j}{m_j}
  \equiv 0
  \label{finP}\,,
  \\
 \sum^N_{j=1}\,  \frac{\p H}{\p \bfp_j}\cdot \overset\leftrightarrow{\mathbb{T}}_{ij} &=
  \sum^N_{j=1}\,  \frac{\bfp_j}{m_j} \cdot \overset\leftrightarrow{\mathbb{T}}_{ij}
  \equiv 0  \,,
  \label{finT}
\\
 \sum^N_{j=1}\, \overset\leftrightarrow{\mathbb{A}}_ {ij}  \cdot \frac{\p H}{\p \bfp_j}
 &= -\sum^N_{j=1}\, \overset\leftrightarrow{\mathbb{A}}_ {ji}  \cdot \frac{\p H}{\p \bfp_j} \equiv 0\,,
 \label{finA}
 \eal
 when evaluated on the constraint $\dot{C}^{A,B}=0$, while  
 \bq
 \sum^N_{j=1}\,  \frac{\p H}{\p \bfp_j}\cdot \overset\leftrightarrow{\mathbb{P}}_{\perp\, ji} \neq 0
  \quad \mathrm{and} \quad  
  \sum^N_{i=1}\,  \frac{\p H}{\p \bfp_i}\cdot \overset\leftrightarrow{\mathbb{T}}_{ij}  \neq 0\,, 
  \label{finPTneq}
 \eq
  when evaluated on the constraint $\dot{C}^{A,B}=0$.
Thus, the bracket of \eqref{geoBkt} yields the equations of motion
 \bal
\dot{\bfq}_i&=\{\bfq_i,H\}_*=\frac{\p H}{\p \bfp_i}=\frac{\bfp_i}{m_i}\,,
\label{eomqrg}\\
\dot{\bfp}_{i}&
=   -  \frac{\p C^A}{\p \bfq_i} \mathbb{S}^{-1}_{AB}  
\sum_{j,k=1}^N \frac{\bfp_j}{m_j}\cdot\frac{\p^2 C^B}{\p \bfq_j\bfq_k} \cdot\frac{\bfp_k}{m_k} \,,
\label{eomprg}
\eal
or
\bq
\ddot{\bfq}_i=- \frac1{m_i} \frac{\p C^A}{\p \bfq_i}
\mathbb{S}^{-1}_{AB}  
\sum_{j,k=1}^N \dot{\bfq}_j \cdot\frac{\p^2 C^B}{\p \bfq_j\bfq_k} \cdot \dot{\bfq}_k \,
= - \sum_{j,k=1}^N\dot{\bfq}_j\cdot \widehat{\boldsymbol{\Ga}}_{i,jk}\cdot \dot{\bfq}_k \,,
\label{finalD}
\eq
where
\bq
 \widehat{\boldsymbol{\Ga}}_{i,jk}:=\frac1{m_i}\frac{\p C^A}{\p \bfq_i}\otimes
\mathbb{S}^{-1}_{AB}  \, 
  \frac{\p^2 C^B}{\p \bfq_j\bfq_k}   \,,
\label{PCS}
\eq
is used to represent the normal force.   

Observe, \eqref{finalD} has two essential features:  as noted, its righthand side is a  normal force that projects to the constraint manifold defined by the constraints $C^A$ and within the constraint manifold it describes a geodesic flow, all done in terms of  the  original Euclidean space coordinates where the  initial conditions place the flow on $\calq$ by setting the values  $C^A$ for all $A=1,2,\dots, M$. We will show this explicitly.

First,  because the components of vectors normal to $\calq$ are given by $\p C^A/\p \bfq_i$ for   $A=1,2,\dots, M$,  this prefactor on the righthand side of \eqref{finalD} projects as expected.  Upon  comparing \eqref{finalD} with \eqref{ELgeod} we  conclude that the coefficient of this prefactor must be  the  Lagrange multipliers, i.e.,  
\bq
\la_A= - \mathbb{S}^{-1}_{AB}  
\sum_{k,j=1}^N \dot{\bfq}_j \cdot\frac{\p^2 C^B}{\p \bfq_j\bfq_k} \cdot \dot{\bfq}_k
\,.
\eq
Thus, we see that Dirac's procedure  explicitly solves  for the Lagrange multiplier. 

Second, to uncover the geodesic flow we can proceed as usual by projecting explicitly onto $\calq$.  To this end
we consider the transformation between the Euclidean configuration space $\E^{3N}$ coordinates
\bq
(\bfq_1, \bfq_2, \dots,\bfq_i,\dots, \bfq_N) \,, \qquad\mathrm{where} \qquad i=1,2,\dots, N 
\eq 
and another set of coordinates 
\bq
(q^1,q^2, \dots,q^a,\dots,q^{3N}) \,,  \qquad\mathrm{where} \qquad a=1,2,\dots, 3N\,, 
\label{untailored}
\eq 
which we  tailor  as follows:
\bq
(q^1,q^2, \dots,q^\al\dots, q^n, C^1,C^2,\dots, C^A,\dots, C^{M})\,,
\eq
where $\al=1,2,\dots, n$, $A=1,2,\dots, M$,  and $n= 3N-M$.    Here we have chosen $q^{n+A}=C^A$ and $n$ is the actual number of degrees of freedom on the constraint surface $\calq$.  We can freely transform back and forth between the two coordinates, i.e.
\bq
(\bfq_1, \bfq_2, \dots,\bfq_i,\dots, \bfq_N) \longleftrightarrow (q^1,q^2, \dots,q^a,\dots,q^{3N}) \,.
\eq
Note, the choice $q^{n+A}=C^A$ could be replaced by $q^{n+A}=f^A(C^1,C^2, \dots, C^M)$ for arbitrary independent $f^A$, but we   assume the original $C^A$ are optimal.   Because $q^\al$ are coordinates within $\calq$, tangent vectors to $\calq$ have the components $\p \bfq_i/\p q^\al$, and there is one for each $\al =1,2,\dots, n$.  The pairing of the normals with  tangents is expressed by
\bq
\sum_{i=1}^N \frac{\p \bfq_i}{\p q^\al}\cdot \frac{\p C^A}{\p \bfq_i} =0\,,
\qquad \al=1,2,\dots,n;\  A=1,2,\dots, M\,.
\eq
 Let us now consider an alternative procedure that the Dirac constraint method provides. Proceeding directly we calculate 
\bq
\dot{q}^a= \sum_{i=1}^N \frac{\p q^a}{\p \bfq_i}\cdot \dot{\bfq}_i\,.
\eq
Observe that on $\E^{3N}$  the matrix ${\p q^a}/{\p \bfq_i}$  is invertible and the full  metric tensor and its inverse in the new coordinates are given as follows:
\bq
g^{ab}=\sum_{i=1}^N\frac1{m_i} \frac{\p q^a}{\p \bfq_i}\cdot\frac{\p q^b}{\p \bfq_i}
\qquad\mathrm{and}\qquad
g_{ab}=\sum_{i=1}^N {m_i} \frac{\p  \bfq_i}{\p q^a}\cdot  \frac{\p  \bfq_i}{\p q^b}\,. 
\label{totalg}
\eq
The metric tensor on $\calq$ of \eqref{gdef}  is obtained by restricting $g_{ab}$  to $a,b\leq n$ and  $g^{\al\be}$ is obtained by inverting  $g_{\al\be}$ and not by restricting $g_{ab}$.
 Proceeding by differentiating again we obtain 
\bq
\ddot{q}^a= \sum_{i=1}^N \frac{\p q^a}{\p \bfq_i}\cdot \ddot{\bfq}_i + 
\sum_{i,j=1}^N \dot{\bfq}_i\cdot \frac{\p^2 q^a}{\p \bfq_i \p  \bfq_j}\cdot \dot{\bfq}_j\,, \qquad a=1,2,\dots, 3N.
\label{ddqa}
\eq
Now inserting  \eqref{finalD} into \eqref{ddqa} gives 
\bq
\ddot{q}^a= - \sum_{i=1}^N  \frac1{m_i}\frac{\p q^a}{\p \bfq_i}\cdot   \frac{\p C^A}{\p \bfq_i}
\, g_{AB}  
\sum_{j,k=1}^N \dot{\bfq}_j \cdot\frac{\p^2 C^B}{\p \bfq_j\bfq_k} \cdot \dot{\bfq}_k
+ 
\sum_{i,j=1}^N \dot{\bfq}_j\cdot \frac{\p^2 q^a}{\p \bfq_i \p  \bfq_j}\cdot  \dot{\bfq}_i\,,  
\label{ddqaa}
\eq
where we have recognized that 
\bq
 g^{AB}= \mathbb{S}^{AB} = 
\sum_{i=1}^N\, \frac1{m_i} 
  \frac{\p C^A}{\p \bfq_i} \cdot \frac{\p C^B}{\p \bfq_i}
  \eq
and, as was necessary for the workability of the Dirac bracket constraint theory,  $g_{AB}= \mathbb{S}^{-1}_{AB}$ must exist.  This quantity is obtained by inverting $\mathbb{S}^{AB}$ and not by restricting $g^{ab}$.

Equation \eqref{ddqaa} is an expression for the full system on $\E^{3N}$.   However,   for $a>n$, we know 
$\ddot{q}^a=\ddot{C}^A=0$, so the  two terms of \eqref{ddqaa} should cancel.   To see this, in the first term of \eqref{ddqaa}  we  set $q^a=C^C$ and observe that this first term  becomes
\bal
 - \sum_{i=1}^N  \frac1{m_i}\frac{\p C^C}{\p \bfq_i}\cdot   \frac{\p C^A}{\p \bfq_i}
\, g_{AB}  
\sum_{j,k=1}^N \dot{\bfq}_j \cdot\frac{\p^2 C^B}{\p \bfq_j\bfq_k} \cdot \dot{\bfq}_k
&=  - g^{CA}\, g_{AB}  
\sum_{j,k=1}^N \dot{\bfq}_j \cdot\frac{\p^2 C^B}{\p \bfq_j\bfq_k} \cdot \dot{\bfq}_k
\nonumber\\
&= - \sum_{j,k=1}^N \dot{\bfq}_j \cdot\frac{\p^2 C^C}{\p \bfq_j\bfq_k} \cdot \dot{\bfq}_k\,. 
\eal
Now,  for $a\leq n$, say $\al$,  the righthand side gives a  Christoffel symbol expression for the geodesic flow; viz. 
\bal
\ddot{q}^\al&= - \sum_{i=1}^N  \frac1{m_i}\frac{\p q^\al}{\p \bfq_i}\cdot   \frac{\p C^A}{\p \bfq_i}
\, g_{AB}  
\sum_{j,k=1}^N \dot{\bfq}_j \cdot\frac{\p^2 C^B}{\p \bfq_j\bfq_k} \cdot \dot{\bfq}_k
+ 
\sum_{j,k=1}^N \dot{\bfq}_j\cdot \frac{\p^2 q^\al}{\p \bfq_j \p  \bfq_k}\cdot \dot{\bfq}_k  
\nonumber \\
&=-\Ga^\al_{\mu\nu}\, \dot{q}^\mu \dot{q}^\nu
\,, 
\label{ddqaa2}
\eal
where
\bq
\Ga^\al_{\mu\nu} =
 \sum_{i=1}^N  \frac1{m_i}\frac{\p q^\al}{\p \bfq_i}\cdot   \frac{\p C^A}{\p \bfq_i}
\, \mathbb{S}^{-1}_{AB}  
\sum_{j,k=1}^N \frac{\p {\bfq}_j}{\p q^\mu}  \cdot\frac{\p^2 C^B}{\p \bfq_j\bfq_k} \cdot \frac{\p {\bfq}_k}{\p q^\nu}
 +  
\sum_{j,k=1}^N \frac{\p {\bfq}_j}{\p q^\mu}  
\cdot \frac{\p^2 q^\al}{\p \bfq_j \p  \bfq_k}\cdot \frac{\p {\bfq}_k}{\p q^\nu}   
\label{newGA}
\eq
is an expression for the Christoffel symbol in terms of the original Euclidean coordinates, the constraints, and the choice of coordinates on $\calq$. 

Using \eqref{newGA} one can calculate an analogous expression for the Riemann curvature tensor on $\calq$ from the usual expression
\bq
R^\al_{\be\ga\de} =  \frac{\p \Ga^\al_{\de\be}}{\p  q^\ga}     -  \frac{\p \Ga^\al_{\ga\be}}{\p  q^\de} 
+ \Ga^\al_{\ga \la} \Ga^\la_{\de \be}  -  \Ga^\al_{\de \la} \Ga^\la_{\ga\be}  \,,
\eq
using $\p /\p q^\ga= \sum_i ( \p \bfq_i/\p q^\ga)\cdot \p /\p \bfq_i$.  This gives the curvature written in terms of the original Euclidean coordinates, the constraints, and the chosen coordinates on $\calq$.

\subsection{$d+1$ field theory}
\label{ssec:d+1}

 The techniques of Sections \ref{ssec:lagrange}, \ref{ssec:dirac},  and \ref{ssec:holoD} have natural extensions to  field theory. 
 
Given independent field variables $\Psi^\cala(\mu,t)$, indexed by $\cala= 1, 2,\dots, \ell$, where the independent variable $\mu =(\mu^1,\mu^2,\dots, \mu^d)$.  The field theoretic version of Hamilton's principle of  \eqref{HamPrin} is embodied in the action 
\bq
 S[\Psi]=\int_{t_0}^{t_1}\!\!dt\, L[\Psi,\dot\Psi] \,,
\qquad \mathrm{with}\qquad 
L[\Psi,\dot\Psi]=\int \! d^d\!\mu\,  \mathcal{L}(\Psi,\dot\Psi,\p \Psi)\,,
\label{fieldLag}
\eq
where we leave the domain of $\mu$ and the boundary conditions unspecified,  but freely drop surface terms obtained upon integration by parts.  The Lagrangian density $\mathcal{L}$  is assumed to depend on the field components $\Psi$ and $\p\Psi$, which is used to indicate all possible partial derivatives with respect of the components of $\mu$.  Hamilton's principle with \eqref{fieldLag} gives the Euler-Lagrange  equations,  
 \bq
 \frac{\de S[\Psi]}{\de\Psi^\cala(\mu,t)} = 0\quad \Rightarrow\quad  \frac{d }{dt} \frac{\p L}{\p \dot{\Psi}^\cala}  +  \frac{\p }{\p \mu} \frac{\p L}{\p \p{\Psi}^\cala} - \frac{\p L}{\p \Psi^\cala}=0\,, 
 \label{FLagEOM}
 \eq
where the overdot implies differentiation at constant $\mu$.   Local holonomic constraints $C^A(\Psi,\p \Psi)$ are enforced  by Lagrange's method by amending the Lagrangian
\bq
L_\la[\Psi,\dot\Psi]=\int \! d^d\!\mu\, \big( \mathcal{L}(\Psi,\dot\Psi,\p \Psi) + \la_A C^A(\Psi,\p \Psi)\big) \,,
\label{CfieldLag}
\eq
with again $A=1,2,\dots, M$ and proceeding as in the finite-dimensional case. 

In the Hamiltonian field theoretic setting, we  could introduce the conjugate momentum densities  $\pi_\cala$,  $\cala= 1, 2,\dots, \ell$,  with the phase space action  
\begin{equation}
S_{\!\la}[\Psi,\pi]=\int_{t_0}^{t_1}\!\!dt\!\!\int \! \!d^d\!\mu\,  
 \left[  \pi_\cala \dot{\Psi}^\cala -\calh + \lambda_A 
D^A\right]\,,   \end{equation}
with Hamiltonian density $\calh$ and local constraints $D^a$ depending on the values of the fields and their conjugates.  Instead of following this route we will jump directly to a generalization of the field theoretic Dirac bracket formalism that would  result.

Consider a Poisson algebra composed of  functionals of field variables ${\chi}^\cala({\mu},t)$   with a Poisson bracket of the form
\begin{equation}
\label{eqn:PBgene}
\{F,G\}=\int\! \!d^d \mu \, F_{\chi} \cdot \mathbb{J}  ({\chi}) \cdot G_{\chi}\,,
\end{equation}
where $F_{\chi}$ is  a shorthand for the functional derivative of a functional $F$ with respect to the field $\chi$ \citep[see e.g.][]{pjm98}  and $F_{\chi} \cdot {\mathbb J} \cdot G_{\chi}= F_{\chi^\cala}  \,{\mathbb J}^{\cala\calb}  \, G_{\chi^\calb}$,  again with repeated indices summed.  Observe the fields 
${\chi}^\cala({\mu},t)$  need not separate into coordinates and momenta, but if they do the Poisson operator  $ \mathbb{J} $ has a form akin to that of \eqref{Jc}. By a Poisson algebra we mean a Lie algebra realization on functionals, meaning the Poisson bracket is bilinear, antisymmetric, and satisfies the Jacobi identity and that there is  an associative product of functionals that satisfies the Leibniz law.  From the Poisson bracket the  equations of motion are given by $\dot{\chi}=\{{\chi},H\}$, for some Hamiltonian functional  $H[{\chi}]$.

 Dirac's constraint theory is generally implemented in terms of canonical Poisson brackets \citep[see e.g.][]{dirac50,sudarshan,sundermeyer},  but it is not difficult to show that his procedure also works for noncanonical Poisson brackets \citep[see e.g.\  an Appendix of][] {pjmLB09}.

We impose an even number of local constraints which we write as $D^a(\mu)=$ constant, a shorthand for $D^a[\chi(\mu)]$, with the index $a=1,2,\dots, 2M$, bearing in mind that they depend on the fields $\chi$ and their derivatives.  As in the finite-dimensional case, the  Dirac bracket is obtained from the matrix $\D$ obtained from the bracket of  the constraints,
\bq
\D^{ab}(\mu,\mu')=\{D^a(\mu),D^b(\mu')\}\,,  
\nonumber
\eq
where we note that $\D^{ab}(\mu,\mu')=-\D^{ba}(\mu',\mu)$.
If $\D$ has an inverse, then the Dirac bracket is defined as follows:
\begin{equation}
\label{eq:dbkt}
\{F,G\}_*=\{F,G\}-\int \!\!d^d\mu\!\! \int\!\! d^d\mu'\,  \{F,D^a(\mu)\}\D^{-1}_{ab}(\mu,\mu')\{D^b(\mu'),G\}\,,
\end{equation}
where  the coefficients $\D^{-1}_{ab}(\mu,\mu')$ satisfy 
$$
\int \!\!d^d\mu' \, \D^{-1}_{ab}(\mu,\mu')\D^{bc}(\mu',\mu'')=\int \!\! d^3\mu' \, 
\D^{cb}(\mu,\mu')\D^{-1}_{ba}(\mu',\mu'')
=\delta_{a}^{c}\delta(\mu-\mu'')\,,
$$
consistent with $\D^{-1}_{ab}(\mu,\mu')=-\D^{-1}_{ba}(\mu',\mu)$.

We note, the procedure is effective only when the coefficients $\D^{-1}_{a b}(\mu,\mu')$ can be found. If $\D$ is not invertible, then one needs, in general, secondary constraints to determine the Dirac bracket.

\subsubsection{Field theoretic  geodesic flow}

In light of Section \ref{sssec:HG},  any field theory with a Lagrangian density of the form
\bq
\call = \frac12 \dot{\Psi}^\cala(\mu,t)\, \eta_{\cala  \calb} \,   \dot{\Psi}^\calb(\mu,t) \,,
\label{FTGlag}
\eq
with the metric $\eta_{\cala  \calb} =\de_{\cala  \calb}$ being the Kronecker delta, subject to time-independent holonomic constraints can be viewed as geodesic flow on the constraint surface.  This is a natural infinite-dimensional generalization of the idea of Section \ref{sssec:HG}.

\section{Unconstrained Hamiltonian and action for fluid}
\label{sec:unconstrained}

\subsection{Fluid action in Lagrangian variable description}
\label{ssec:fluidaction}

 The Lagrangian variable description of a fluid is described in Lagrange's famous work \citep{lagrange}, while historic and additional material can be found in  \citet{serrin59,newcomb62,VKF67,pjm98}.   Because  the Lagrangian description treats a fluid as a continuum of particles, it naturally is amenable to the Hamiltonian form.  The {Lagrangian variable} is a  coordinate that gives the position of a fluid element or parcel,  as it is sometimes called,   at time $t$.  We  denote this variable  by  $\bfq=\bfq(\bfa,t)=(q^1,q^2,q^3)$, which is measured relative to some cartesian coordinate system.    
Here  $\bfa=(a^1,a^2,a^3)$ denotes the {fluid element label}, which is often  defined to be  the position of the fluid element at the initial  time, $\bfa=\bfq(\bfa,0)$, but this need not always be the case.  The label $\bfa$ is a continuum analog of  the discrete index that labels a generalized coordinate in a finite degree-of-freedom system.   If $D$ is a   domain that is fully occupied by the fluid, then at each fixed time $t$,  $\bfq\colon D\rightarrow D$  is assumed to be 1-1 and onto. Not much is really known about the mathematical properties of this function, but we will assume that it is as smooth as it needs to be for the operations performed.  Also, we will assume we can freely integrate by parts dropping surface terms and drop reference to D in our integrals. 

When discussing the ideal fluid and MHD we will use repeated sum notation with upper and lower indices even though we are working in cartesian coordinates.  And, unlike in Section 2,  latin indices, $i,j,k,\ell$ etc.\ will be summed over 1,2, and 3, the cartesian components,  rather than to $N$.  This is done to avoid further proliferation of indices and we trust  confusion will not arise because of context.

Important quantities are  the deformation matrix, $\p q^i/\p a^j$ and its Jacobian  determinant $\mathcal{J}:= \det(\p q^i/\p a^j)$,  which   is given by
\bq
\mathcal{J}= \frac1{6}\ep_{kj\ell}\ep^{imn}  \frac{\p q^k}{\p a^i}  \frac{\p q^j}{\p a^m} \frac{\p q^\ell}{\p a^n}  \,,
\label{J3}
\eq
where  $\ep_{ijk}=\ep^{ijk}$  is the  purely antisymmetric (Levi-Civita) tensor density. 
We assume a fluid element is uniquely determined by its label for all time.  Thus,  $\mathcal{J}\neq 0$  and   we can invert $\bfq=\bfq(\bfa,t)$ to obtain the label associated with the fluid element at position 
$\bfx$ at time $t$, $\bfa=\bfq^{-1}(\bfx,t)$.  
For coordinate transformations $\bfq=\bfq(\bfa,t)$ we have 
\bq
\frac{\p q^{k}}{\p a^j}\, \frac{A^i_k}{\calj} = \de^i_j\,,  
\label{inverseAJ}
\eq
where $A^i_k$ is the cofactor of $\p q^k/\p a^i$ , which can be written as follows:
\bq
A^i_k= \frac12\ep_{kj\ell}\ep^{imn} \frac{\p q^{j}}{\p a^m}\frac{\p q^{\ell}}{\p a^n}\,.
 \label{co3}
\eq

 Using  $\bfq(\bfa,t)$ or its inverse $\bfq^{-1}(\bfx,t)$, various quantities  can be written either as a function of $\bfx$ or $\bfa$.  For convenience we list additional formulas below for latter use:
 \bal
\mathcal{J} & =  \frac{1}{3}A_{\ell}^{k}\frac{\partial q^{\ell}}{\partial a^{k}}\,,
\\
 A^j_i &=  \frac{\p \mathcal{J}}{\p (\p q^i/\p a^j)} \,,
 \\
\frac{\partial (A_{i}^{k}f)}{\partial a^{k}}&=A_{i}^{k}\,\frac{\partial f}{\partial a^{k}}\,,
\label{dAda}
\\
\de \calj &=  A_{i}^{k}\frac{\partial \de{q}^{i}}{\partial a^{k}} \hspace{1.33cm} \mathrm{or}\qquad \dot{\calj}=A_{i}^{k}\frac{\partial \dot{q}^{i}}{\partial a^{k}}\,,
\label{dedet}
\\
\delta\left(\frac{A_{\ell}^{k}}{\mathcal{J}}\right)\frac{\partial q^{\ell}}{\partial a^{u}}&=-\frac{A_{i}^{k}}{\mathcal{J}}\frac{\partial}{\partial a^{u}}\delta q^{i}
\qquad \mathrm{or}\qquad
\delta\left(\frac{A_{\ell}^{k}}{\mathcal{J}}\right)=-\frac{A_{i}^{k}A_{\ell}^{u}}{\mathcal{J}^{2}}\frac{\partial}{\partial a^{u}}\delta q^{i}\,, 
\label{eq:AJvar}
\\
 {{A}_{\ell}^{u} \frac{\partial}{\partial a^{u}}\left[\frac{A_{i}^{k}}{\calj} \frac{\partial f}{\partial a^{k}} \right]}
    & {=A_{i}^{k}\frac{\partial}{\partial a^{k}}\left[\frac{A_{\ell}^{u}}{\calj}\frac{\partial f}{\partial a^{u}}\right]\,,\forall\, f}
\,,
\label{IDen}
\eal
which follow from  the standard rule for differentiation of determinants and the expression for the cofactor matrix.   {For example, the commutator expression of \eqref{IDen} follows easily from \eqref{eq:AJvar}, which in turn follows upon differentiating \eqref{inverseAJ}.  These formulas are all of classical origin, e.g., the second equation of  \eqref{dedet} is the Lagrangian variable version of a  formula due  to Euler \citep[see e.g.][] {serrin59}. }

Now we are in a position to recreate and generalize Lagrange's Lagrangian for the ideal fluid action principle.  On physical grounds we expect our fluid to possess kinetic and  internal energies, and if magnetized, a  magnetic energy.  The total kinetic energy functional of the fluid is naturally given by
\bq
T[\dot{\bfq}]:=\frac1{2} \int \!d^3a\, \rho_0(\bfa)\,  |\dot \bfq|^2\,,
\label{Tq}
\eq 
where $\rho_0$ is the mass density attached to the fluid element labeled by $\bfa$ and $\dot\bfq$ denotes time differentiation  of $\bfq$ at fixed label $\bfa$. Note, in \eqref{Tq}  
$|\dot \bfq|^2=\dot{q}_i\dot{q}^i$, where in general $\dot{q}_i=g_{ij}\,\dot{q}^i$, but we will only consider the cartesian metric where $g_{ij}=\de_{ij}=\eta_{ij}$.

Fluids are assumed to be in local thermodynamic equilibrium and thus can be  described by a function $U(\rh,s)$, an internal energy per unit mass that depends on the specific volume $\rho^{-1}$ and specific entropy $s$.  If a magnetic field $\bfB(\bfx,t)$   were present, then we could add dependence on $|\bfB|$ as in \citet{pjm82}  to account for pressure anisotropy.  { \citep[See also][where this appears  in  the context of gyroviscosity.]{pjmLA14,pjmLW20}} The internal energy  is written in terms of the Eulerian density and entropy (see Section \ref{ssec:LtoE}) since we expect the fluid at each Eulerian observation point to be in thermal equilibrium.    From $U$ we compute  the temperature and pressure according to the usual differentiations, $T=\p U/\p s$ and $p=\rh^2\p U/\p {\rh}$.   For MHD, the magnetic energy    $H_B=\int\! d^3x\,  |\bfB|^2/2$ in Lagrangian variables  would be added.   For the ideal fluid,  the total internal energy functional is
 \bq
 V[\bfq]:=  \int \!d^3a\, \rho_0 \, U\left({\rho_0}/{\calj}, s_0\right)
 \,.
\label{Vq}
 \eq
Here we have used the fact that a fluid element carries a specfic entropy $s=s_0(\bfa)$ and a mass determined by $\rho=\rho_0(\bfa)/\calj$.   In Section \ref{ssec:LtoE} we will describe in detail  the map from Lagrangian to Eulerian variables.

Thus, the special case of the action principle of \eqref{fieldLag} for the ideal fluid has  Lagrange's Lagrangian $L[\bfq,\dot\bfq]=T - V$.  Variation of this action gives the Lagrangian equation of motion for the fluid
\bq
\rho_0 \ddot{q}_i=- A^j_i\,   \frac{\p}{\p a^j} 
\left(
\frac{\rho_0^2}{\calj^2} \frac{\p U}{\p \rho}
\right)\,, 
\label{lagEOM}
\eq
with an additional term that describes the $\bfJ\times\bfB$ force in Lagrangian variables for MHD.  See, e.g.,  \citet{newcomb62,pjm98,pjm09} for details of this calculation and the MHD extension. 

\subsection{Hamiltonian formalism in  Lagrangian description}
\label{ssec:HamDesc}

Upon defining the momentum density as usual by
\bq
\pi_i=\frac{\de L}{\de \dot{q}^i}=\rho_0 \,\dot{q}_i\,, 
\label{piDef}
\eq
we can obtain the Hamiltonian by Legendre transformation, yielding
\bq
H[{\bfpi},\bfq]=T + V=  \int \!d^3a\,\left( \frac{|{\bfpi}|^2}{2   \rho_0 }\, 
+  \rho_0   U\left({\rho_0}/{\calj}, s_0\right)\right)\,, 
\label{LagHam}
\eq
where $|\bfpi|^2= \pi^i\pi_i= \pi_i \eta^{ij} \pi_j$.
This Hamiltonian with the canonical Poisson bracket, 
\begin{equation}
\left\{F,G\right\} =\int\! d^{3}a\,\left(\frac{\delta F}{\delta q^{i}}\frac{\delta G}{\delta\pi_{i}}-\frac{\delta G}{\delta q^{i}}\frac{\delta F}{\delta\pi_{i}}\right)\,,
\label{cbkt}
\end{equation}
yields 
\bal
\dot{q}^i&= \{q^i, H\}= \pi^i\!/\rho_0  \,,
\label{qdot}
\\
\dot{\pi}_i&= \{\pi_i, H\}= -  
A^j_i\,  \frac{\p}{\p a^j} 
\left(
\frac{\rho_0^2}{\calj^2} \frac{\p U}{\p \rho}
\right)\,.
\label{pidot1}
\eal
Equations \eqref{qdot} and \eqref{pidot1} are equivalent to \eqref{lagEOM}.   For MHD a term $H_B$ is added to \eqref{LagHam}   \citep[see][] {newcomb62,pjm09}.  We will give this explicitly in the constraint context in Section \ref{sssec:LDholo} after discussing the Lagrange to Euler map. 

\subsection{Hamiltonian formalism in  Eulerian description via the Lagrange to Euler map}
\label{ssec:LtoE}

In order to understand how constraints in terms of the Lagrangian variable description relate to those in terms of the Eulerian description, in particular  $\nabla\cdot \bfv=0$, it is necessary to understand the mapping from Lagrangian to Eulerian variables.  
Thus, we record here the relationship between the two  unconstrained descriptions, i.e., how the noncanonical Hamiltonian structure of the compressible Euler's equations relates  to the Hamiltonian structure described in Section \ref{ssec:HamDesc}.

For the ideal fluid, the set of Eulerian variables can be taken to be $\{\bfv,\rho, s\}$, where $\bfv(\bfx,t)$ is the velocity field at the Eulerian observation point,  $\bfx=(x,y,z)=(x^1,x^2,x^3)$ at time $t$ and,  as as noted in Section  \ref{ssec:fluidaction},  $\rho(\bfx,t)$ is the mass density and  $s(\bfx,t)$ is the specific entropy. 
In order to describe magnetofluids  the magnetic field $\bfB(\bfx,t)$  would be appended to this set.  It is most important to distinguish between the Lagrangian fluid element position and label variables, 
$\bfq$ and $\bfa$,  and the Eulerian observation point $\bfx$, the latter two being independent   variables.  Confusion exists in the literature because some authors use  the same symbol for   the Lagrangian coordinate $\bfq$ and the Eulerian observation point $\bfx$.

The Lagrangian and Eulerian descriptions must clearly be related and, indeed,   knowing $\bfq(\bfa,t)$ we can obtain $\bfv(\bfx,t)$.  If one were to insert a velocity probe into a fluid at $(\bfx,t)$ then one would measure the velocity of the fluid element that happened to be at that position at that time.   Thus it is clear that $\dot{\bfq}(\bfa,t)=\bfv(\bfx,t)$, where recall the  overdot means the time derivative  at constant $\bfa$.  But,  which fluid element will be at $\bfx$ at time $t$?  Evidently $\bfx=\bfq(\bfa,t)$, which upon inversion yields the label of that element that will be measured, 
$\bfa=\bfq^{-1}(\bfx,t)$.  Thus, the Eulerian velocity field is given by 
\bq
\bfv(\bfx,t) =\left.\dot{\bfq}(\bfa,t)\right|_{\bfa=\bfq^{-1}(\bfx,t)}=\dot{\bfq}\circ\bfq^{-1}(\bfx,t)\,.
\eq
Properties can be attached to fluid elements, just as a given mass is identified with a given particle in mechanics. For a continuum system it is natural to attach a mass density, $\rho_0(\bfa)$, to the element labeled by $\bfa$.  Whence the element of mass in a given volume is given by $\rho_0d^3a$ and this amount of mass is preserved by the flow, i.e.\ $\rho(\bfx,t)d^3x=\rho_0(\bfa)d^3a$.  Because the locus of points of material surfaces  move with the fluid are  determined by $\bfq$, an initial volume element $d^3a$ maps into a volume element $d^3x$ at time $t$  according to  
\bq
d^3x=\mathcal{J} d^3a\,,
\label{vol}
\eq
Thus,  using \eqref{vol} we obtain $\rho_0=\rho \calj$ as used in Section \ref{ssec:fluidaction}.

  Other quantities could be attached to a fluid element;  for the ideal fluid,  entropy per unit mass, $s(\bfx,t)$,  is such a quantity.  The assumption that each fluid element is isentropic then amounts to $s=s_0$.  Similarly, for MHD a magentic field, $B_0(\bfa)$,  can be attached, and then the frozen flux assumption  yields $B\cdot d^2x=B_0\cdot d^2a$. An initial area element $d^2a$ maps into an area  element $d^2x$ at time $t$  according to  
\bq
(d^2x)_i=  A^j_i \, (d^2a)_j\,.
\label{area}
\eq 
Using  (\ref{area}) we obtain $\calj B^i=B_0^j\, \p q^i/\p a^j$.

Sometimes it is  convenient to use  another set of Eulerian density variables: $\{\bfM,\rho, \si,\bfB\}$, where $\si=\rh s$  is the entropy per unit volume, and $\bfM=\rho \bfv$ is the momentum density.   These  Eulerian variables  can be represented by using the Dirac delta function to `pluck out' the fluid element that happens to be at the Eulerian observation point $\bfx$ at time $t$. For example, the  mass  density $\rh(\bfx,t)$ is obtained by
\bq
\rh({\bfx},t)=\int \!d^3a
\, \rh_0(\bfa) \, \de\left({\bfx}-{\bfq}\left(\bfa,t\right)\right)
=\left. \frac{\rh_0}{\mathcal{J}}\right|_{\bfa=\bfq^{-1}({\bfx},t)}\,.
\label{rhoEu}
\eq
The density one observes at $\bfx$ at time $t$ will be the one attached to the fluid element that happens to be there then, and this fluid element has a label given by solving $\bfx=\bfq(\bfa,t)$.  The second equality  of (\ref{rhoEu}) is obtained by using the three-dimensional version of the delta function identity $\de(f(x))=\sum_i \de (x-x_i)/|f'(x_i)|$, where $f(x_i)=0$.  Similarly,  the entropy per unit volume is given by 
\bq
\si({\bfx},t)=\int\! d^3a\, 
\si_0(\bfa) \, \de\left({\bfx}-{\bfq}\left(\bfa,t\right)\right) 
=\left. \frac{\si_0}{\mathcal{J}}\right|_{\bfa=\bfq^{-1}({\bfx},t)}\,,
\label{siEu}
\eq
which is consistent with $\si_0(\bfa)=\rh_0(\bfa) s_0(\bfa)$ and $s(\bfx,t)=\left. s_0(\bfa)\right|_{\bfa=\bfq^{-1}({\bfx},t)}$, where the latter means $s$ is constant along a Lagrangian orbit.  Proceeding, the  momentum density, $\bfM=(M_1,M_2,M_3)$, is  related to the Lagrangian canonical momentum (defined in Section \ref{ssec:HamDesc}) by  
\bq
\bfM(\bfx,t)=\int  \!d^3a \, 
\bfpi(\bfa,t) \, \de\left({\bfx}-{\bfq}(\bfa,t)\right) 
= \left.
\frac{{\bfpi}(\bfa,t)}{\mathcal{J}}
\right|_{\bfa=\bfq^{-1}(\bfx,t)}\,, 
\label{MEu}
\eq
where for the ideal fluid and MHD, \  ${\bfpi}(\bfa,t)=(\pi_1,\pi_2,\pi_3)=\rh_0 \dot{\bfq}$.  Lastly, 
\bq
B^i(\bfx,t)=\int \!d^3a \, \frac{\p q^i(\bfa,t)}{\p a^j}  B_0^j(\bfa) \, 
\de\left({\bfx}-{\bfq}(\bfa,t)\right)  
= \left.
\frac{\p q^i(\bfa,t)}{\p a^j}\frac{B_0^j(\bfa)}{\mathcal{J}} 
\right|_{\bfa=\bfq^{-1}(\bfx,t)}\,,
\eq
for the components of the magnetic field.  It may be  unfamiliar to view the magnetic field as  density, but in MHD  it obeys a conservation law.   {Geometrically, however, it naturally satisfies the equations of a vector density associated with a differential 2-form as was observed in  \citet{pjm82} and \citet{TY93}. }

To obtain the noncanonical  Eulerian Poisson bracket we consider functionals $F[\bfq,\mathbf{\pi}]$ that are restricted so as to obtain their dependence on $\bfq$ and $\mathbf{\pi}$ only through the Eulerian variables. Upon setting $F[\bfq,\bfpi]=\bar{F}[\bfv,\rho,\si]$,  equating variations of both sides,
\bq
 \delta F = \int d^3a \left[\frac{\delta F}{\delta \bfq}\cdot\delta \bfq +
 \frac{\delta F}{\delta \mathbf{\pi}}\cdot\delta \mathbf{\pi}\right] 
 = \int d^3x \,\left[\frac{\delta\bar F}{\delta\rho}\delta\rho +
 \frac{\delta\bar F}{\delta \sigma}\delta \sigma +
\frac{\delta \bar F}{\delta \bfM}
 \cdot\delta  \bfM\right]=\de\bar{F}\,,
 \label{deltaF}
 \eq
varying the  expressions \eqref{rhoEu}, \eqref{siEu}, and \eqref{MEu}, substituting the result into 
\eqref{deltaF}, and equating the independent coefficients of $\de\bfq$ and $\de{\bfpi}$, we obtain
\bal
\frac{\delta F}{\delta \bfq}& = \int d^3x\,\left[\rho_ 0\,\nabla
\frac{\delta\bar F}{\delta\rho} + \sigma_0\,\nabla
  \frac{\delta\bar F}{\delta\sigma}
 + \pi_{i} \nabla \frac{\delta\bar F}{\delta M_{i}} \right]\,
\delta(\bfx-\bfq)  \,,
 \label{deFq}
 \\
 \frac{\delta F}{\delta {\bfpi}} &= \int d^3x\, \frac{\delta\bar F}{\delta \bfM} \,
\delta (\bfx-\bfq)\,.
\label{deFpi}
\eal
(See \cite{pjm98} and \cite{pjmG80} for details.)
 Upon substitution of   \eqref{deFq} and   \eqref{deFpi}, expressions of the functional chain rule that relate Lagrangian functional derivatives to the Eulerian functional derivates,  into \eqref{cbkt}
yields  the  following bracket expressed entirely in terms of the Eulerian fields $\{\bfM,\rho,\si\}$: 
\begin{eqnarray}
 \{F,G\} &=& -\int \!d^3 x\, \Bigg[
 M_i \Bigg(\frac{\delta F}{\delta M_j} \frac{\partial}{\partial x^j}
\frac{\delta G}{\delta M_i}  - \frac{\delta G}{\delta M_j}
\frac{\partial}{\partial x^j} \frac{\delta F}{\delta M_i} \Bigg)
\nonumber\\
 &+& \rho \, \Bigg(\frac{\delta F}{\delta \bfM} \cdot
\nabla \frac{\delta G}{\delta \rho}
 - \frac{\delta G}{\delta \bfM} \cdot \nabla
\frac{\delta F}{\delta \rho}\Bigg)
+  \sigma\,  \Bigg(\frac{\delta F}{\delta  \bfM} \cdot
\nabla \frac{\delta G}{\delta \sigma} -
 \frac{\delta G}{\delta  \bfM} \cdot \nabla
\frac{\delta F}{\delta \sigma}\Bigg)
\Bigg] \,.
\label{Mbkt} 
 \end{eqnarray}
 In \eqref{Mbkt}  we have dropped the overbars on the Eulerian functional derivatives.
The bracket for MHD is the above with the addition of the following term, which is obtained by adding a $\bfB$ contribution to \eqref{deltaF}:
\bal      
   \{F,G\}_B &=-\int \!d^3 x\, \Bigg[ \bfB\cdot
    \left(
      \frac{\delta F}{\delta \bfM} \cdot\nabla \frac{\delta G}{\delta \bfB}
-     \frac{\delta G}{\delta \bfM} \cdot\nabla \frac{\delta F}{\delta \bfB}
\right)
\nonumber\\
&\hspace{1cm} +  \bfB\cdot
    \left(
 \nabla \left( \frac{\delta F}{\delta \bfM} \right ) \cdot \frac{\delta G}{\delta \bfB}
-    \nabla \left( \frac{\delta G}{\delta \bfM} \right ) \cdot \frac{\delta F}{\delta \bfB}
\right)
 \Bigg]\,,  
 \label{Bbkt}
 \eal
where dyadic notation is used;  for example, $\bfB\cdot[\nabla (\bfD) \cdot \bfC]=\sum_{i,j} B_i C_j \p D_j/\p x_i$,  for vectors $\bfB,\bfD$, and $\bfC$. Alternatively, the bracket in terms of 
$\{\bfv,\rho,s,\bfB\}$ is obtained using chain rule expressions, e.g., 
\bq
 \frac{\delta F}{\delta \rho}\Bigg|_{\bfv,s} = \frac{\delta F}{\delta
\rho}\Bigg|_{\bfM,s}
 +\frac{\bfM}{\rho}\cdot \frac{\delta F}{\delta \bfM} + \frac{\sigma}{\rho}
\frac{\delta F}{\delta \sigma}\,, 
 \label{chn1}
 \eq
 yielding
\bal
 \{F,G\} &= -\int\!d^3 x \, \Bigg[\Bigg(
    \frac{\delta F}{\delta \rho}\nabla\cdot\frac{\delta G}{\delta \bfv}
   - \frac{\delta G}{\delta \rho}\nabla\cdot\frac{\delta F}{\delta \bfv}\Bigg)
 + \Bigg(\frac{\nabla \times \bfv}{\rho}\cdot\frac{\delta G}{\delta
 \bfv}\times\frac{\delta F}{\delta \bfv}\Bigg) 
 \nonumber\\
 &\hspace{2cm} + \frac{\nabla s}{\rho}\cdot\Bigg(\frac{\delta F}{\delta s}
\frac{\delta G}{\delta \bfv} -
            \frac{\delta G}{\delta s}\frac{\delta F}{\delta \bfv}\Bigg)
            \Bigg]
\,,
 \label{vbkt}
\eal
and
\bal
 \{F,G\}_B & =  -\int\!d^3 x \, \Bigg[
 \bfB\cdot
    \left(
    \frac{1}{\rho} \frac{\delta F}{\delta \bfv} \cdot\nabla \frac{\delta G}{\delta \bfB}
-    \frac{1}{\rho} \frac{\delta G}{\delta \bfv} \cdot\nabla \frac{\delta F}{\delta \bfB}
\right)
\nonumber\\
&\hspace{1cm}
+\bfB\cdot
    \left(
 \nabla \left(\frac{1}{\rho} \frac{\delta F}{\delta \bfv} \right ) \cdot \frac{\delta G}{\delta \bfB}
-    \nabla \left(\frac{1}{\rho} \frac{\delta G}{\delta \bfv} \right ) \cdot \frac{\delta F}{\delta \bfB}
\right)
            \Bigg]\,.
 \label{vBbkt}
\eal

 The bracket of \eqref{vbkt} plus that of \eqref{vBbkt}  with the Hamiltonian
\bq
H[\rho, s,\bfv,\bfB]=\int\! d^3x\left(\frac12 \rho |\bfv|^2 +\rho U(\rho,s) +\frac12 |\bfB|^2\right)
\eq
gives the Eulerian version of MHD in Hamiltonian form, $\p \bfv/\p t=\{\bfv,H\}$, etc., and similarly using  \eqref{Mbkt} plus \eqref{Bbkt} with the Hamiltonian expressed in terms of $(\bfM,\rho, \si,\bfB)$.  Ideal fluid follows upon neglecting the $\bfB$ terms.

\subsection{Constants of motion:  Eulerian vs.\ Lagrangian}
\label{ssec:CoM}

In oder to compare the imposition of constraints in the Lagrangian and Eulerian descriptions, it is necessary to compare Lagrangian and Eulerian conservations laws.  This is because constraints, when enforced, are  conserved quantities.  The comparison is not trivial  because time independent quantities in the Eulerian description can be time dependent in the Lagrangian description. 

Consider a Lagrangian  function $f(\bfa,t)$, typical of the Lagrangian variable description, and  
the relation $\bfx=\bfq(\bfa,t)$, which relates an Eulerian observation point $\bfx$ to a corresponding fluid element trajectory value. The function $f$ can be written in either picture by composition, as follows:
\bq
f(\bfa,t)=\tilde{f}(\bfx,t)= \tilde{f}(\bfq(\bfa,t),t)\,,
\eq
where we will use a tilde to indicated the Eulerian form of a Lagrangian function. Application of  the chain rule gives 
\bq
\left. \frac{A^i_k}{\calj} \, \frac{\p f}{\p a^i}\,\right|_{\bfa=\bfq^{-1}(\bfx,t)} =\frac{\p \tilde{f}}{\p x^k} 
\qquad \mathrm{and}\qquad
 \left. \frac{A_{\ell}^{k}}{\calj}\,\frac{\partial}{\partial a^{k}}\left(\frac{\pi^{\ell}}{\mathcal{J}}\right)\right|_{\bfa=\bfq^{-1}(\bfx,t)} =  \nabla\cdot \bfv \,,
\label{ELgrad}
\eq
with the second equality of \eqref{ELgrad} being a special case of the first. Similarly, 
\bq
 \left.\dot f \, \right|_{\bfa=\bfq^{-1}(\bfx,t)}=  \frac{\p \tilde{f}}{\p t} + \left.  \dot{q}^i(\bfa,t)\, \frac{\p \tilde{f}}{\p x^i}\,\right|_{\bfa=\bfq^{-1}(\bfx,t)}= 
 \frac{\p \tilde{f}}{\p t} + \bfv\cdot \nabla   \tilde{f}(\bfx,t)\,, 
 \label{lagDet}
\eq
where recall an overdot denotes the  time derivative at constant $\bfa$,     $\p/\p t$   denotes the  time derivative at constant $\bfx$, and $\nabla$ is the Eulerian gradient with components $\p/\p x^i$ as used in \eqref{ELgrad}.   Because the Jacobian determinant $\calj$ is composed of derivatives of $\bfq$,  we have $\calj(\bfa,t)|_{\bfa=\bfq^{-1}(\bfx,t)} =\tilde{\calj}(\bfx,t)$, whence one obtains a   formula due  to Euler \citep[see e.g.][] {serrin59}, 
\bq
\frac{\p \tilde\calj}{\p t} + \bfv\cdot \nabla \tilde \calj= \tilde\calj \, \na\cdot \bfv\,,
\label{Jeuler}
\eq
 {which can be compared to its Lagrangian version of \eqref{dedet}.}

Now, consider a    conservation law in the Lagrangian variable description,
\bq
\dot \cald_L + \frac{\p \Ga_{\cald_L}^i}{\p a^{i}}=0\,,
\label{Lcon}
\eq
where  the density $\cald_L(\bfa,t)$ has the associated flux  $\bfGa_{\cald_L}$.   Then, the  associated conserved  quantity  is 
\bq
\cali_{\cald_L}=\int \!d^3a \, \cald_L\,, 
\label{ILcon}
\eq
which satisfies $d\cali_{\cald_L}/dt =0$ provided surface terms vanish.  Similarly, an Eulerian conservation law with density $\cald_E$ and flux ${\bfGa}_{\cald_E}$ is
\bq
\frac{\p\cald_E}{\p t} + \frac{\p  \Ga_{\cald_E}^i}{\p x^{i}}=0
\label{Econ}
\eq
and the following is similarly constant in time:
\bq
\cali_{\cald_E}=\int \!d^3x \,  \cald_E\,. 
\eq
The relationship between the two conservation laws (\ref{Lcon}) and (\ref{Econ}) can be obtained by defining
\bq
\tilde\cald_L={\calj}\cald_E\,,   \qquad \tilde\Ga^i_{\cald_L}= A^i_k\, \bar\Ga_{\cald_E}^k\,,
\qquad {\rm and}\qquad  \mathbf{\Ga}_{\cald_E}= \mathbf{ \bar\Ga}_{\cald_E} + \bfv\, \cald_E\,,
\label{ELcon}
\eq
and their equivalence follows from (\ref{dedet}), (\ref{lagDet}), and (\ref{Jeuler}).  Given a Lagrangian conservation law, one can use \eqref{ELcon} to obtain a corresponding Eulerian conservation law.  The density $\cald_E$ is obtained from the first equation of \eqref{ELcon}, a piece of the Eulerian flux $\bar{\bfGa}_{\cald_E}$ from the second, which then can be substitued into the third equation of \eqref{ELcon} to obtain the complete Eulerian flux $\bfGa_{\cald_E}$.   An Eulerian conservation law is most useful when one can write $\cald_E$ and $\bfGa_{\cald_E}$ entirely  in terms of the Eulerian variables of the fluid.

The simplest case occurs when $\cald_L$ only depends on $\bfa$, in which case the  corresponding flux is zero and 
$\p  \cald_L/\p t=0$ and $d\cali_{\cald_L}/dt=0$ follow directly because (\ref{ILcon}) has no time dependence whatsoever.  Any attribute attached to a fluid element only depends on the label $\bfa$ and this has a trivial conservation law of this form.  However, such trivial Lagrangian conservation laws yield nontrivial Eulerian conservation laws.  Observe, even thought 
$\mathbf{\bar\Ga}_{\cald_E}\equiv 0$ by \eqref{ELcon},  $\bfGa_{\cald_E}= \bfv \cald_E\neq 0$.  Consider the case of the entropy where $\cald_L=s_0(\bfa)$,  whence  $s(\bfx,t)=s_0(\bfa(\bfx,t))$ and by  \eqref{lagDet}, 
\bq
\frac{\p s}{\p t} +\bfv\cdot \na s=0\,,
\eq
with the quantity $s=s_0/\calj$ being according to  \eqref{ELcon} the Eulerian conserved density, as can be verified using \eqref{Jeuler}. But,  as it stands, this density cannot be written in terms of Eulerian fluid variables.  However, $\si_0=\rho_0s_0$ is also a trivial Lagrangian conserved density and  according to (\ref{ELcon})  we have the Eulerian density $\rho_0s_0/\calj=\rho s=\si$ that satisfies
\bq
\frac{\p \si}{\p t} +  \na \cdot(\bfv\si)=0\,.
\eq
Thus, it follows that any advected scalar has an associated conserved quantity obtained by multiplication by  $\rh$.  

As another example, consider the  quantity $B_0^i \p q^j/\p a^i$.  This quantity is the limit   displacement  between two nearby fluid elements, i.e.,  $\bfq(\bfa,t) -\bfq(\bfa +\de\bfa,t)$ along the initial magnetic field as $\de \bfa \rightarrow 0$.  Evidently,
\bq
\dot{\left(B_0^i\frac{ \p q^j}{\p a^i}\right)}= B_0^i\frac{ \p \dot{q}^j}{\p a^i}
= \frac{\p}{\p a^i} \left(B_0^i \dot{q}^j\right)\,,
\label{Bpara}
\eq
where the second equality follows if the initial  magnetic field is divergence free. This is of course another trivial conservation law, for the time derivative of a density that is a divergence will always be a divergence.  However,  let us see what this becomes in the Eulerian description.  According to \eqref{ELcon} the corresponding Eulerian density is  $\cald_E=\cald_L/\calj$; so,  the density associated with this trivial conservation law \eqref{Bpara} is 
\bq
B^j(\bfx,t) = \left.\frac{B_0^i}{\calj}\frac{ \p q^j}{\p a^i}\right|_{\bfa=\bfq^{-1}(\bfx,t)}\,.
\eq
which as we saw in Section \ref{ssec:LtoE} is the expression one gets for the MHD magnetic field because of  flux conservation.  That the divergence-free magnetic field satisfies a conservation law is clear from 
\bq
\frac{\p \bfB}{\p t}= -\bfv\cdot \nabla \bfB+ \bfB\cdot \nabla \bfv - \bfB\, \nabla \cdot \bfv=\nabla \cdot \overset\leftrightarrow{{T}}\,,
\label{Bcon}
\eq
where the tensor $\overset\leftrightarrow{{T}}$ of the last equality is 
\bq
\overset\leftrightarrow{{T}} = \bfB \otimes\bfv-\bfv \otimes \bfB\,.
\eq
Thus we have another instance where a trivial Lagrangian conservation law leads to a nontrivial  Eulerian one.

Although  $B_0^i { \p q^j}/{\p a^i}$ does not map into an expression entirely in terms of our set of Eulerian variables, we can divide it by $\rho_0$, a quantity that only depends on the label $\bfa$, and  obtain
\bq  
\left.\frac{B_0^i}{\rho_0}\frac{ \p q^j}{\p a^i}\right|_{\bfa=\bfq^{-1}(\bfx,t)}= \frac{B^j}{\rho}\,.
\eq
 Eulerianizing the counterpart of \eqref{Bpara} for this expression gives
 \bq
 \frac{\p }{\p t} \left(\frac{\bfB}{\rho}\right) + \bfv\cdot \nabla \left(\frac{\bfB}{\rho}\right)
 =  \frac{\bfB}{\rho} \cdot \nabla \bfv\,, 
 \label{Brho}
 \eq
which is  not an Eulerian  conservation law.  This is to be expected because, unlike what we did to get   \eqref{Bcon},  we have Eulerianized without using  \eqref{ELcon}.   In light of its relationship to $\bfq(\bfa,t) -\bfq(\bfa +\de\bfa,t)$, the quantity $\bfB/\rho$ has been described as a measure of the distance of points on a magnetic field line \citep[see e.g.][] {vkampen}.  This was predated by analogous arguments for vorticity \citep[see e.g.][] {serrin59}.

\section{Constraint theories for the incompressible ideal fluid}
\label{sec:dirac}

\subsection{The incompressible fluid in  Lagrangian  variables}
\label{ssec:LagCon}

In order to enforce incompressibility, Lagrange added to his Lagrangian the constraint $\calj=1$ with the Lagrange multiplier $\la(\bfa,t)$,
\bq
L_\la[\bfq,\dot{\bfq}]= T[\dot{\bfq}]  + \la \, \calj\,, 
\label{lagla}
\eq
with $T$ given \eqref{Tq}.  Here  we have dropped $V$ because incompressible fluids contain no internal energy.   Upon insertion of \eqref{lagla} into the action of Hamilton's principle it is discovered that $\la$ corresponds to the pressure.  The essence of this procedure was known to Lagrange.    (See \citet{serrin59}  for historical details and \citet{sommerfeld} for an elementary exposition.)  This procedure yields
\bq
\rho_0\ddot q^i=-A^i_j \, \frac{\p \la}{\p a^j}\,,
\label{lageomla}
\eq
where use has been made of \eqref{dedet}.  The Eulerian form of \eqref{lageomla} is clearly 
$\rho(\p \bfv/\p t+\bfv\cdot\nabla \bfv)=-\nabla \la$, whence it is clear that $\la$ is the pressure.  Although  Lagrange  knew the  Lagrange multiplier was the pressure, he could only solve for it in special cases.  The general procedure of Section \ref{ssec:bgnd} was not available because Green's function techniques and the theory of elliptic equations were not at his disposal.

\subsubsection{Lagrangian volume preserving geodesic flow}

 If the constraint is dropped from \eqref{lagla}, we obtain free particle motion for an infinite-dimensional system, the ideal fluid case of \eqref{FTGlag} of Section \ref{ssec:d+1}, which is analogous to  the finite-dimensional case  of Section \ref{sssec:HG}. Because the constraint 
 $\calj=1$ only depends on the derivatives of $\bfq$, it is a configuration space constraint;  thus,  it   is an  holonomic constraint.   As is well-known and reviewed in Section \ref{sssec:HG}, free particle motion with holonomic constaints is  geodesic flow.  Thus, following Lagrange, it is immediate that    the ideal incompressible fluid is an infinite-dimensional version of geodesic flow.

Lagrange's calculation was placed in a geometric/group theoretic  setting  in \cite{arnold-diffeo} (see also Appendix 2 of \cite{arnold-book} and \cite{khesin}). Given that the transformation  $\bfa\leftrightarrow \bfq$, at any time, is assumed to be a smooth invertible coordinate change, it is a Lie group, one  referred to as the diffeomorphism group.  With the additional assumption that these transformations are volume preserving,  Lagrange's constraint  $\calj=1$, the transformations form a subgroup, the group of volume preserving diffeomorphisms.  Thus, Lagrange's work can be viewed as geodesic flow on the group of volume preserving diffeomorphisms. 

Although Arnold's assumptions of smoothness etc.\ are mathematically dramatic, his description of Lagrange's calculations in these terms  has spawned a considerable body of research.  Associated with a geodesic flow is a metric, and whence one can calculate a curvature. In his original work, Arnold added the novel calculation of the curvature in the mathematically more forgiving case of two-dimensional flow with periodic boundary conditions.

\subsection{Lagrangian-Dirac constraint theory}
\label{ssec:LDconTh}

More recently there have been several works \citep{pjmTC09,pjmCGBT13,pjmLB09},
following \citet{turski99,turski01}, that treat  the enforcement of
the incompressibility constraint of hydrodynamics by Dirac's method
of constraints \citep{dirac50}. In these works the compressibility
constraint was enforced in the Eulerian variable description
of the fluid using the noncanonical Poisson bracket of Section \ref{ssec:LtoE} 
as the base bracket of a generalization of Dirac's constraint theory. We will return to this approach in Section \ref{ssec:LtoE} where we revisit and extend Dirac's  constraint results for the fluid in the Eulerian variable description.  Here, apparently for the first time,  we consider the incompressibility constraint in the Lagrangian variable description, where the canonical Poisson bracket  of \eqref{cbkt} is the base for the construction of a Dirac bracket. 

We adapt \eqref{eq:dbkt} for the fluid case at hand with the supposition of only two local  constraints, 
which we write as
\bq
D^a(\bfa') =\int d^3a \,D^a(\bfa)\, \de(\bfa-\bfa')\,,
\label{deexp}
\eq
where $a=1,2$ and $D^a(\bfa)$ is a shorthand for a function of $\bfq(\bfa, t)$ and $\bfpi(\bfa,t)$ and their derivatives with respect to $\bfa$.  Then the matrix $\D$ is a $2\times 2$ matrix with the components
\bq
\D^{ab}(\bfa,\bfa')=\{D^a(\bfa),D^b(\bfa')\}\,,  
\nonumber
\eq
 using the canonical bracket of  \eqref{cbkt}.
To construct the Dirac bracket 
\begin{equation}
\label{eqn:DBF}
\{F,G\}_*=\{F,G\}-\int \!\!d^3a\!\! \int\!\! d^3a'\,  \{F,D^a(\bfa)\}\D^{-1}_{ab}(\bfa,\bfa')\{D^b(\bfa'),G\},
\end{equation}
we require the inverse,  which satisfies
\begin{equation}
\int\!d^{3}a\,\,\D^{ac}(\mathbf{a}^{\prime},\mathbf{a})\,\D^{-1}_{cb}(\mathbf{a},\mathbf{a}^{\prime\prime})=\delta^{a}_{b}\,\delta(\mathbf{a}^{\prime}-\mathbf{a}^{\prime\prime})\,.
\label{inverse}
\end{equation}

Rather than continuing  with the general case, which is unwieldy, we proceed to the special case for the incompressible fluid, an infinite-dimensional version of the holonomic constraints  discussed in Section \ref{sssec:HCD}.

\subsubsection{Lagrangian-Dirac incompressibility  holonomic constraint}
\label{sssec:LDholo}

Evidently we will want our holonomic incompressibility constraint to be $\calj$.  However, it is convenient to express this by choosing
\bq
D^1=\ln\left(\frac{\mathcal{J}}{\rho_{0}}\right)\,.
\label{D1}
\eq
This amounts to the same constraint as $\calj=1$ with the value $D^1=-\ln(\rho_0)$.   The scaling of $\calj$ in \eqref{D1} by  $\rho_0(\bfa)$ is immaterial because it is a time-independent quantity.  
 To obtain the second constraint we follow suit and  set 
\begin{equation}
 D^{2}=\dot{D}^1=\frac{A_{\ell}^{k}}{\mathcal{J}}\,\frac{\partial}{\partial a^{k}}\left(\frac{\pi^{\ell}}{\rho_{0}}\right)
 = \eta^{\ell j}\,\frac{A_{\ell}^{k}}{\mathcal{J}}\,\frac{\partial}{\partial a^{k}}\left(\frac{\pi_{j}}{\rho_{0}}\right)\,,
 \label{constraints-1-1}
\end{equation}
where recall we assume $\eta^{\ell j}=\de^{\ell j}$ and $\pi_j$ is given by \eqref{piDef}.  That the  constraint $D^2$ is the time derivative of $D^1$ requires the definition of $\pi_j$ of \eqref{piDef} that uses the Hamiltonian $ \int \!d^3a\  {|{\bfpi}|^2}/{(2\rho_0)}$.

Observe, that  constraints $D^1$ and $D^2$  are local constraints in that they are
enforced pointwise \citep[see e.g.][] {pjmF11}, i.e., they are enforced
on each fluid element labeled by $\mathbf{a}$. Equation (\ref{D1})
corresponds in the Eulerian picture  to $-\ln(\rho)$, while 
the second constraint
of (\ref{constraints-1-1}), the Lagrangian  time derivative  of the
first constraint,  corresponds in the Eulerian picture  to $\nabla\cdot\mathbf{v}$, 
which can be easily verified using the second equation of \eqref{ELgrad}.
Note, the particular values of these constraints of interest are, of course,
$\mathcal{J}=1$ and $\nabla\cdot\mathbf{v}= 0$, but the dynamics
the Dirac bracket generates will preserve any values of these constraints.
For example, we could set $\mathcal{J}=f\left(\mathbf{a}\right)$
where the arbitrary function $f$ is less than unity for some $\mathbf{a}$
and greater for others, corresponding to regions of fluid elements
that experience contraction and expansion.  Also note, because we have used $\bfpi$ with the up index in \eqref{constraints-1-1}; thus as seen in the second equality it depends on the metric.  This was done to make it have the Eulerian form 
$\nabla\cdot\mathbf{v}$.

For the constraints \eqref{D1} and \eqref{constraints-1-1},  $\D$  only depends on two  quantities because  $D^{1}$ does not depend on $\bfpi$, i.e.\  $\{D^{1},D^{1}\}= 0$ and $\{D^{1},D^{2}\}=-\{D^{2},D^{1}\}$.
Thus the inverse has the form 
\begin{equation}
\D^{-1}=\left(\begin{matrix}\D^{-1}_{11} & \D^{-1}_{12}\\
\D^{-1}_{21} & 0
\end{matrix}\right)\,,
\end{equation}
giving rise to the conditions 
\begin{equation}
\D^{-1}_{12}\cdot \D^{21}=\mathcal{I}=\D^{-1}_{21}\cdot \D^{12}\, \qquad
\mathrm{and}
\qquad 
\D^{-1}_{11}\cdot \D^{12}+ \D^{-1}_{12}\cdot \D^{22}=0
\,,
\label{condition}
\end{equation}
where $\mathcal{I}$ is the identity. Thus, the inverse is easily tractable if the inverse of $\D^{12}$
exists; whence, 
\begin{equation}
\D^{-1}_{11}=-\D^{-1}_{12}\cdot \D^{22}\cdot \D^{-1}_{21}\,.
\label{condition2}
\end{equation}
In the above the symbol `$\,\cdot\,$' is used to denote the product
with the sum in infinite dimensions, i.e., integration over $d^3a$ as in \eqref{inverse}.
Equation (\ref{condition2}) can be rewritten in an abbreviated form with implied integrals on repeated arguments as
\begin{equation}
\D^{-1}_{11}(\mathbf{a}^{\prime},\mathbf{a}^{\prime\prime})=\D^{-1}_{21}(\mathbf{a}^{\prime},\hat{\mathbf{a}})\cdot \D^{22}(\hat{\mathbf{a}},\check{\mathbf{a}}) 
\cdot\D^{-1}_{21}(\check{\mathbf{a}},\mathbf{a}^{\prime\prime})\,.
\label{condition2-1}
\end{equation}

In order to obtain $\D$ and its inverse, we need  the functional derivatives of $D^{1}$ and $D^2$.  These are obtained directly by writing these local constraints as in \eqref{inverse}, yielding 
\bal
\frac{\delta D^{1}(\mathbf{a}^{\prime})}{\delta q^{i}(\mathbf{a})} & =  -A_{i}^{k}\frac{\partial}{\partial a^{k}}\frac{\delta(\mathbf{a}-\mathbf{a}^{\prime})}{\mathcal{J}} \,,
\label{D1q}\\
\frac{\delta D^{1}(\mathbf{a}^{\prime})}{\delta\pi_{i}(\mathbf{a})} & =  0
\label{D1pi}
\,,
\eal
where use has been made of \eqref{dedet}, and 
\bal
\frac{\delta D^{2}(\mathbf{a}^{\prime})}{\delta q^{i}(\mathbf{a})} & =  \frac{\partial}{\partial a^{u}}\left(\frac{A_{i}^{k}A_{\ell}^{u}}{\mathcal{J}^{2}}\frac{\partial}{\partial a^{k}}\left(\frac{\pi^{\ell}}{\rho_{0}}\right)\delta(\mathbf{a}-\mathbf{a}^{\prime})\right) \,,
\label{D2pi}\\
\frac{\delta D^{2}\left(\mathbf{a}^{\prime}\right)}{\delta\pi_{i}(\mathbf{a})} & =  -\frac{\, \eta^{ij}}{\rho_{0}}\frac{\partial}{\partial a^{m}}\left(\frac{A_{j}^{m}}{\mathcal{J}}\delta\left(\mathbf{a}-\mathbf{a}^{\prime}\right)\right)\,,
\label{D2q}
\eal
where use has been made of  (\ref{eq:AJvar}) and recalling  we have  \eqref{dAda} at our disposal.

Let us now insert \eqref{D1q}, \eqref{D1pi}, \eqref{D2pi}, and  \eqref{D2q}  into the canonical 
Poisson bracket \eqref{cbkt}, to  obtain 
\bal
\D^{12}(\mathbf{a},\mathbf{a}^{\prime})  &=  \{D^{1}(\mathbf{a}),D^{2}(\mathbf{a}^{\prime})\}
\nonumber\\
 &=   -\frac{A_{i}^{\ell}}{\mathcal{J}}\frac{\partial}{\partial a^{\ell}}\left(\frac{\eta^{ij}}{\rho_{0}}
  {A_{j}^{k}}   \frac{\partial}{\partial a^{k}}\left(\frac{\delta(\mathbf{a}-\mathbf{a}^{\prime})}{\mathcal{J}}\right)\right) \,, 
 \label{D12}
 \eal
 which corresponds to the symmetric matrix $\mathbb{S}$ of \eqref{DAB}  and \eqref{fpDAB} and 
 \bal
\D^{22}({\mathbf{a}},\bfa')  &= 
\{D^{2}({\mathbf{a}}),D^{2}(\bfa')\}
\nonumber
\\
& =    \frac{A_{i}^{k}A_{\ell}^{u}}{\mathcal{J}^{2}}\frac{\partial}{\partial a^{k}}\left(\frac{\pi^{\ell}}{\rho_{0}}\right)\frac{\partial}{\partial a^{u}}\left[\frac{ \eta^{ij}}{\rho_{0}}\frac{\partial}{\partial a^{m}}\left(\frac{A_{j}^{m}}{\mathcal{J}}\delta\left(\mathbf{a}-\bfa'\right)\right)\right] 
\nonumber\\
& \hspace{1.5cm}   -  \frac{A_{i}^{m}}{\mathcal{J}}\frac{\partial}{\partial a^{m}}
 \left[\frac{\eta^{ij}}{\rho_{0}}\frac{\partial}{\partial a^{u}}\left(\frac{A_{j}^{k}A_{\ell}^{u}}{\mathcal{J}^{2}}\frac{\partial}{\partial a^{k}}\left(\frac{\pi^{\ell}}{\rho_{0}}\right)\delta(\mathbf{a}-\bfa')
 \right)\right] \,,
\label{eq:C22-2} 
\eal
which corresponds to the antisymmetric matrix $\mathbb{A}$ of  \eqref{D2D2}.  Observe the symmetries corresponding to the matrices $\mathbb{S}$ and $\mathbb{A}$, respectively, are here 
 \bal
 \int \! d^3a' \, \D^{12}(\bfa,\bfa')\, \phi (\bfa')&=\int d^3a' \D^{12}(\bfa',\bfa)\, \phi (\bfa') \,,
\nonumber\\
 \int \! d^3a' \, \D^{22}(\bfa,\bfa')\, \phi (\bfa')&=-\int d^3a' \D^{22}(\bfa',\bfa)\, \phi (\bfa')\,, 
 \nonumber
 \eal
for all functions $\phi$.  The first follows from integration by parts, while the second  is obvious from its definition.

Using (\ref{inverse}), the first condition of  \eqref{condition} is  
\begin{equation}
\int\!d^{3}a^{\prime\prime}\,\,\D^{12}(\mathbf{a}^{\prime},\mathbf{a^{\prime\prime}})\,\D^{-1}_{21}(\mathbf{a^{\prime\prime}},\hat{\mathbf{a}})=\delta(\mathbf{a}^{\prime}-\hat{\mathbf{a}})\,,
\label{inverseA}
\end{equation}
which upon substitution of  \eqref{D12} and integration gives
\begin{equation}
-\frac{A_{i}^{\ell}}{\calj} \frac{\partial}{\partial a^{\ell}} 
\left[ \frac{\eta^{ij}}{\rho_{0}} A_{j}^{k}\, \frac{\partial}{\partial a^{k}}\left(\frac{\mathbb{D}_{21}^{-1}(\mathbf{a},\mathbf{a}^{\prime\prime})}{\mathcal{J}}\right)\right]=\delta(\mathbf{a}-\mathbf{a}^{\prime\prime})\,.
\label{eq:ddDinv21}
\end{equation}
 
We introduce  the formally self-adjoint operator (cf.\  \eqref{dAda})
\begin{equation}
\Delta_{\rho_{0}}f:=\frac{A_{i}^{\ell} }{\mathcal{J}}\frac{\partial}{\partial a^{\ell}}\left[ \frac{\eta^{ij}}{\rho_{0}} A_{j}^{k} \frac{\partial}{\partial a^{k}}\left(\frac{f}{\mathcal{J}}\right)\right]\,,
\end{equation}
i.e., an operator that satisfies 
\bq
\int\!d^3a\,  f(\bfa) \, \Delta_{\rho_0} g(\bfa) =\int\!d^3a \, g(\bfa)  \, \Delta_{\rho_0} f(\bfa) \,,
\eq 
a property inherited by its  inverse $\Delta_{\rho_{0}}^{-1}$.   Thus  we can rewrite equation (\ref{eq:ddDinv21}) as
\begin{equation}
\mathbb{D}_{21}^{-1}(\mathbf{a},\mathbf{a}^{\prime\prime}) 
=-G_{0}\left(\mathbf{a},\mathbf{a}^{\prime\prime}\right)=-\Delta_{\rho_{0}}^{-1}\delta(\mathbf{a}-\mathbf{a}^{\prime\prime})\,,
\label{eq:Dinv21-1}
\end{equation}
where $G_{0}$ represents the Green function associated with (\ref{eq:ddDinv21}).

In order to obtain $\D^{-1}_{21}$,  we find it convenient to  transform \eqref{eq:Dinv21-1}  to Eulerian variables.  Using  $\mathbf{x}=\mathbf{q}(\bfa,t)$ we find
\begin{equation}
\frac{\D^{-1}_{21}(\mathbf{a},\mathbf{a}^{\prime})}{\mathcal{J}}=-G(\mathbf{x},\mathbf{x}^{\prime})=-G(\mathbf{q}(\mathbf{a}),\mathbf{q}(\mathbf{a}^{\prime}))\,, 
\end{equation}
where   $G$ satisfies
\begin{equation}
\nabla\cdot\left(\frac{1}{\rho}\nabla G\right)=-\Delta_{\rho}\frac{\D^{-1}_{21}(\mathbf{a},\mathbf{a}^{\prime})}{\mathcal{J}}=\mathcal{J}\delta(\mathbf{x}-\mathbf{x}^{\prime}).
\end{equation}
Here use has been made of identities  \eqref{dAda} and \eqref{ELgrad}.  
As noted in Section \ref{ssec:bgnd}, under physically reasonable conditions, the operator 
\bq
\Delta_{\rho}f=\Delta_{\rho_{0}}\left(\mathcal{J}f\right)=\nabla\cdot\left(\frac1{\rho} \nabla f\right)
\label{nabrho}
\eq
 has an inverse.  Thus we write 
\begin{equation}
 \D^{-1}_{21}(\mathbf{a},\mathbf{a}^{\prime})=-\calj \Delta_{\rho}^{-1}
 \Big(\mathcal{J}\,
 \delta\big(\mathbf{q}(\bfa,t)-\mathbf{q}(\bfa',t)\big)
 \Big)\,. 
 \label{eq:Dinv21}
\end{equation}
Now, using $\D^{-1}_{21}=-\D^{-1}_{12}$, the element $\D^{-1}_{11}$ follows directly from (\ref{condition2}).

For convenience we write the  Dirac bracket  of \eqref{eqn:DBF} as follows:
\begin{equation}
\left\{ F,G\right\} _{*}=\left\{ F,G\right\} -\left[F,G\right]^{D} \,,
\label{eq:dbkt1}
\end{equation}
where
\begin{align}
\left[F,G\right]^{D}& :=\sum_{a,b=1}^{2}\left[ F,G\right]^D_{ab}=\int\!d^{3}a\,\int\!d^{3}a^{\prime}\,\,\left\{ F,D^{a}\left(\mathbf{a}\right)\right\} \mathbb{D}_{ab}^{-1}(\mathbf{a},\mathbf{a}^{\prime})\left\{ D^{b}\left(\mathbf{a}^{\prime}\right),G\right\} \,.
\label{eq:FG}
\end{align}
Because  $\mathbb{D}_{22}^{-1}=0$ and $\left[ F,G\right]^D _{12}=-[ G,F]^D_{21}$, we only need to calculate  $[F,G]^D_{11}$
and $[F,G]^D_{21}$. 

As above, we  substitute  \eqref{D1q}, \eqref{D1pi}, \eqref{D2pi}, and \eqref{D2q}  into the bracket (\ref{cbkt}) 
and  obtain 
\begin{align}
\left\{ F,D^{1}\left(\mathbf{a}\right)\right\}  & =-\frac{A_{i}^{k}}{\calj} \frac{\partial}{\partial a^{k}}\left(\frac{\delta F}{\delta\pi_{i}}\right),\label{eq:F-D1}\\
\left\{ F,D^{2}\left(\mathbf{a}\right)\right\}  & = \frac{A_{\ell}^{k}}{\calj} 
\frac{\partial}{\partial a^{k}}\left(\frac{\eta^{i\ell } }{\rho_{0}}\frac{\delta F}{\delta q^{i}}\right) 
+\frac{A_{i}^{k}}{\calj} \frac{A_{\ell}^{u}}{\calj} \frac{\partial}{\partial a^{k}}\left(\frac{\pi^{\ell}}{\rho_{0}}\right)\,\frac{\partial}{\partial a^{u}}\left(\frac{\delta F}{\delta\pi_{i}}\right)\,.
\label{eq:F-D2}
\end{align}
Then,  exploiting the antisymmetry of the Poisson bracket, it is straightforward
to calculate analogous expressions for the terms $\left\{ D^{1,2},G\right\}$. 

We first analyze the operator 
\begin{align}
[F,G]^D_{11} & =\!\int\!d^{3}a\!\!\int\!\!d^{3}a^{\prime}\!\!\int\!d^{3}\hat{a}\!\!\int\!d^{3}\check{a} \left\{ F,D^{1}\!\left(\mathbf{a}\right)\right\} \mathbb{D}_{21}^{-1}(\mathbf{a},\hat{\mathbf{a}})\,\mathbb{D}^{22}(\hat{\mathbf{a}},\check{\mathbf{a}})\,\mathbb{D}_{21}^{-1}(\check{\mathbf{a}},\mathbf{a}^{\prime})\left\{ D^{1}\!\left(\mathbf{a}^{\prime}\right),G\right\} ,
\end{align}
where we used  the second condition of \eqref{condition} to replace $\mathbb{D}_{11}^{-1}$.
Upon inserting   \eqref{eq:Dinv21-1} and  \eqref{eq:F-D1}, this equation can
be rewritten as
\begin{align}
[F,G]^D_{11} & =-\int\!d^{3}a\,\int\!d^{3}a^{\prime}\,\int\!d^{3}\hat{a}\,\,\int\!d^{3}\check{a} 
\nonumber \\
 &\hspace{1cm}  \left[\frac{A^h_{j}}{\calj}\frac{\partial}{\partial a^{h}}\left(\frac{\delta F}{\delta\pi_{j}}\right)\Delta_{\rho_{0}}^{-1}\delta(\mathbf{a}-\hat{\mathbf{a}})\right]_{\bfa=\mathbf{a}}\!\!\!
 \mathbb{D}^{22}(\hat{\mathbf{a}},\check{\mathbf{a}})\,\left[\frac{A_{r}^{s}}{\calj}\frac{\partial}{\partial a^{s}}\left(\frac{\delta G}{\delta\pi_{r}}\right)\Delta_{\rho_{0}}^{-1}\delta(\check{\mathbf{a}}-\mathbf{a})\right]_{\bfa=\mathbf{a}^{\prime}} , 
\end{align}
where the subscripts on the right delimiters  indicate that  $\mathbf{a}$
is to be replaced after the derivative operations,  including those that occur   in $\mathcal{J}$
and $A_{i}^{j}$.

Integrating  this expression by parts 
with respect to $\mathbf{a}$ and $\mathbf{a}^{\prime}$ yields 
\begin{align}
[F,G]^D_{11} & =-\int\!d^{3}\hat{a}\,\,\int\!d^{3}\check{a}\,\,\left[\Delta_{\rho_{0}}^{-1}\left(\frac{A_{j}^{h}}{\calj}\frac{\partial}{\partial a^{h}}\left(\frac{\delta F}{\delta\pi_{j}}\right)\right)\right]_{\bfa=\hat{\mathbf{a}}}\!\!\!\mathbb{D}^{22}(\hat{\mathbf{a}},\check{\mathbf{a}})\,\left[\Delta_{\rho_{0}}^{-1}\left(\frac{A_{r}^{s}}{\calj}\frac{\partial}{\partial a^{s}}\left(\frac{\delta G}{\delta\pi_{r}}\right)\right)\right]_{\bfa=\check{\mathbf{a}}} ,
\label{eq:FG11}
\end{align}
and then substituting \eqref{eq:C22-2} gives 
\begin{align}
[F,G]^D_{11} & =-\!\int\!d^{3}\hat{a}\!\!\int\!d^{3}\check{a}
\left[\Delta_{\rho_{0}}^{-1}\left(\frac{A_{j}^{h}}{\calj}\frac{\partial}{\partial a^{h}}\left(\frac{\delta F}{\delta\pi_{j}}\right)\right)\right]_{\bfa=\hat{\mathbf{a}}}\!
\left\{\frac{A_{i}^{k} A_{\ell}^{u}}{\calj^2}\frac{\partial}{\partial a^{k}}
\left(\frac{\pi^{\ell}}{\rho_{0}}\right)\frac{\partial}{\partial a^{u}}\left[\frac{\eta^{in}}{\rho_{0}}\frac{\partial}{\partial a^{m}}\left(\frac{A_{n}^{m}}{\calj}\delta\left(\mathbf{a}-\check{\mathbf{a}}\right)\right)\right]\right.\nonumber \\
 & \hspace{.25cm} \left.-  \frac{A_{i}^{m}}{\calj} \frac{\partial}{\partial a^{m}}\left[\frac{\eta^{in}}{\rho_{0}}\frac{\partial}{\partial a^{u}}\left(\frac{A_{n}^{k}}{\calj}\frac{A_{\ell}^{u}}{\calj}\frac{\partial}{\partial a^{k}}\left(\frac{\pi^{\ell}}{\rho_{0}}\right)\delta(\mathbf{a}-\check{\mathbf{a}})\right)\right]
 \right\} _{\bfa=\hat{\mathbf{a}}}\,
 \left[\Delta_{\rho_{0}}^{-1}\left(\frac{A_{r}^{s}}{\calj} \frac{\partial}{\partial a^{s}}\left(\frac{\delta G}{\delta\pi_{r}}\right)\right)\right]_{\bfa=\check{\mathbf{a}}}.
 \label{eq:FG11-bis}
\end{align}
Then, by means of integrations by parts we can remove the derivatives
from the term $\delta(\mathbf{a}-\check{\mathbf{a}})$ and perform
the integral. After relabeling the integration variable as $\mathbf{a}$
to simplify the notation,  (\ref{eq:FG11-bis}) becomes
\begin{align}
[F,G]^D_{11} & =\int\!d^{3}a\,\,  \mathlarger{\Bigg\{}
\rho_{0}\, \eta^{ui}\frac{A_{u}^{k}}{\calj}\frac{\partial}{\partial a^{k}}\left(\frac{\pi^{\ell}}{\rho_{0}}\right)
\left(\left.\mathbb{P}_{\rho_0\perp}\frac{\delta F}{\delta\boldsymbol{\pi}}\right|_{\ell}\left.\mathbb{P}_{\rho_0\perp}\frac{\delta G}{\delta\boldsymbol{\pi}}\right|_{i} 
- \left.\mathbb{P}_{\rho_0\perp}\frac{\delta F}{\delta\boldsymbol{\pi}}\right|_{i}\left. 
\mathbb{P}_{\rho_0\perp}\frac{\delta G}{\delta\boldsymbol{\pi}}\right|_{\ell}\right) 
\nonumber \\
 &\hspace{1cm}  + \, \eta^{ni} {A_{\ell}^{u}}  \frac{\partial}{\partial a^{u}}\left[\frac{A_{n}^{k}}{\calj}\frac{\partial}{\partial a^{k}}\left(\frac{\pi^{\ell}}{\rho_{0}}\right)\right]\left[\left.\mathbb{P}_{\rho_0\perp}\frac{\delta G}{\delta\boldsymbol{\pi}}\right|_{i}\frac{\Delta_{\rho_{0}}^{-1}}{\mathcal{J}}\left(\frac{A_{j}^{h}}{\calj} \frac{\partial}{\partial a^{h}}\left(\frac{\delta F}{\delta\pi_{j}}\right)\right)\right.
 \nonumber \\
 & \hspace{5cm}\left.-\left.\mathbb{P}_{\rho_0\perp}\frac{\delta F}{\delta\boldsymbol{\pi}}\right|_{i}\frac{\Delta_{\rho_{0}}^{-1}}{\mathcal{J}}\left(\frac{A_{j}^{h}}{\calj}\frac{\partial}{\partial a^{h}}\left(\frac{\delta G}{\delta\pi_{j}}\right)\right)\right]
  \mathlarger{\Bigg\}}\,,
 \label{eq:FG11-L}
\end{align}
where we introduced the projection operator
\bq
\left(\mathbb{P}_{\rho_0\perp}\right)^{i}_j\, {z}^j= 
\frac{\eta^{i\ell}}{\rho_{0}} A_{\ell}^{u}  \frac{\partial}{\partial a^{u}}\left[\frac{\Delta_{\rho_{0}}^{-1}}{\mathcal{J}}\left(\frac{A_{j}^{h}}{\calj}\frac{\partial}{\partial a^{h}}\, z^{j}\right)\right] 
=: \mathbb{P}_{\rho_0\perp}\mathbf{z}\,\big|^{i}\,,
\label{eq:P0perp}
\eq
where in the last equality we defined a shorthand for convenience;  {thus,  
\bq
\left.\mathbb{P}_{\rho_0\perp}\frac{\de F}{\de \boldsymbol{\pi}}\,\right|_{\ell}:= \frac1{\rho_{0}} A_{\ell}^{u}  \frac{\partial}{\partial a^{u}}\left[\frac{\Delta_{\rho_{0}}^{-1}}{\mathcal{J}}\left(\frac{A_{j}^{h}}{\calj}\frac{\partial}{\partial a^{h}}\, 
\frac{\de F}{\de \pi_{j}}\right)\right]\,.
\eq
}

It is straightforward to prove that $\mathbb{P}_{\rho_0\perp}$ represents
a projection, i.e.\  $\mathbb{P}_{\rho_0\perp}\left(\mathbb{P}_{\rho_0\perp}\mathbf{z}\right)=\mathbb{P}_{\rho_0\perp}\mathbf{z}$ for each $\mathbf{z}$, which in terms of  indices would have an  $i$th component given by  $(\mathbb{P}_{\rho_0\perp})^i_j (\mathbb{P}_{\rho_0\perp})^j_k \,z^k = (\mathbb{P}_{\rho_0\perp})^i_ k\,  z^k$. 
Also, $\mathbb{P}_{\rho_0\perp}$ is formally self-adjoint with respect to the following weighted inner product:
\bq
\int d^3a \, \rho_0 \, w_i \, (\mathbb{P}_{\rho_0\perp})^i_j \, z^j = \int d^3a \, \rho_0 \, z_i \, (\mathbb{P}_{\rho_0\perp})^i_j \, w^j \,.
\label{PperpSA}
\eq
The projection operator complementary to $\mathbb{P}_{\rho_0\perp}$ is given by
\bq
\mathbb{P}_{\rho_0}=  I - \mathbb{P}_{\rho_0\perp}\,,
\label{P}
\eq
 where $I$ is the identity. 
 
Now let us return to our evaluation of $[ F,G]_D$ and  analyze the contribution  
\begin{align}
[F,G]^D_{21}  & =\int\!d^{3}a\,\int\!d^{3}a^{\prime}\,\,\left\{ F,D^{2}\left(\mathbf{a}\right)\right\} \mathbb{D}_{21}^{-1}(\mathbf{a},\mathbf{a}^{\prime})\left\{ D^{1}\left(\mathbf{a}^{\prime}\right),G\right\} .
\end{align}
Using \eqref{eq:Dinv21},  \eqref{eq:F-D1}, and \eqref{eq:F-D2}, this
equation can be rewritten as 
\begin{align}
[F,G]^D_{21} & =-\int\!d^{3}a\,\int\!d^{3}a^{\prime}\,\,\left[
\frac{A_{i}^{k}}{\calj}\frac{\partial}{\partial a^{k}}\left(\frac{\eta^{in}}{\rho_{0}}\frac{\delta F}{\delta q^{n}}\right)+\frac{A_{i}^{k}}{\calj} \frac{A_{\ell}^{u}}{\calj} \frac{\partial}{\partial a^{k}}\left(\frac{\pi^{\ell}}{\rho_{0}}\right)\,\frac{\partial}{\partial a^{u}}\left(\frac{\delta F}{\delta\pi_{i}}\right)\right]\nonumber \\
 &\hspace{3.5cm} \times   \Delta_{\rho_{0}}^{-1}\delta(\mathbf{a}-\mathbf{a}^{\prime})\left[\frac{A_{j}^{h}}{\calj}\frac{\partial}{\partial a^{h}}\left(\frac{\delta G}{\delta\pi_{j}}\right)\right]_{\bfa=\mathbf{a}^{\prime}}
\end{align}
and, integrating by parts to simplify the $\delta(\mathbf{a}-\mathbf{a}^{\prime})$
term, results in
\begin{align}
[F,G]^D_{21}  & =\int\!d^{3}a\,\, \mathlarger{\Bigg\{}
  \frac{\delta F}{\delta q^{i}}\left.\mathbb{P}_{\rho_0\perp}\frac{\delta G}{\delta\boldsymbol{\pi}}\right|^{i} 
  +\rho_{0}\frac{A_{i}^{k}}{\calj} \frac{\partial}{\partial a^{k}}\left(\frac{\pi^{\ell}}{\rho_{0}}\right)\,\frac{\delta F}{\delta\pi_{i}}\left.\mathbb{P}_{\rho_0\perp}\frac{\delta G}{\delta\boldsymbol{\pi}}\right|_{\ell}\nonumber \\
 &\hspace{1cm}  + A_{\ell}^{u}\frac{\partial}{\partial a^{u}}\left[\frac{A_{i}^{k}}{\calj} \frac{\partial}{\partial a^{k}}\left(\frac{\pi^{\ell}}{\rho_{0}}\right)\right]\,\frac{\delta F}{\delta\pi_{i}}\frac{\Delta_{\rho_{0}}^{-1}}{\mathcal{J}}\left(\frac{A_{j}^{h}}{\calj} \frac{\partial}{\partial a^{h}}\left(\frac{\delta G}{\delta\pi_{j}}\right)\right)
 \mathlarger{\Bigg\}}
 \,.
 \label{eq:FG21-L}
\end{align}

We can now combine the operators $[F,G]^D_{11}$,
$[F,G]^D_{21}$,  and $[F,G]^D_{12}=-[G,F]^D_{21}$,
given by (\ref{eq:FG11-L}) and (\ref{eq:FG21-L}), to calculate the
Dirac bracket (\ref{eq:dbkt1}). 
First,  we  rewrite (\ref{eq:FG}) as 
\begin{align}
[F,G]^D  & =\int\!d^{3}a\,\, \mathlarger{\Bigg\{}
\frac{\delta F}{\delta q^{i}}\left.\mathbb{P}_{\rho_0\perp}\frac{\delta G}{\delta\boldsymbol{\pi}}\right|^{i}-\frac{\delta G}{\delta q^{i}}\left.\mathbb{P}_{\rho_0\perp}\frac{\delta F}{\delta\boldsymbol{\pi}}\right|^{i}
\nonumber \\
 & +\rho_{0} \frac{A_{i}^{k}}{\calj} \frac{\partial}{\partial a^{k}}\left(\frac{\pi^{\ell}}{\rho_{0}}\right)\,
 \left(\left.\mathbb{P}_{\rho_0} \frac{\delta F}{\delta\boldsymbol{\pi}}\right|^{i}\left.\mathbb{P}_{\rho_0\perp}\frac{\delta G}{\delta\boldsymbol{\pi}}\right|_{\ell}-\left.\mathbb{P}_{\rho_0}\frac{\delta G}{\delta\boldsymbol{\pi}}\right|^{i}\left.\mathbb{P}_{\rho_0\perp}\frac{\delta F}{\delta\boldsymbol{\pi}}\right|_{\ell}\right)
 \nonumber \\
 & + A_{\ell}^{u}\frac{\partial}{\partial a^{u}}\left[\frac{A_{i}^{k}}{\calj} \frac{\partial}{\partial a^{k}}\left(\frac{\pi^{\ell}}{\rho_{0}}\right)\right]\,\left[\left.\mathbb{P}_{\rho_0}\frac{\delta F}{\delta\boldsymbol{\pi}}\right|^{i}\frac{\Delta_{\rho_{0}}^{-1}}{\mathcal{J}}\left(\frac{A_{j}^{h}}{\calj} \frac{\partial}{\partial a^{h}}\left(\frac{\delta G}{\delta\pi_{j}}\right)\right)\right.
 \nonumber \\
 & \hspace{4cm}\left.-\left.\mathbb{P}_{\rho_0}\frac{\delta G}{\delta\boldsymbol{\pi}}\right|^{i}\frac{\Delta_{\rho_{0}}^{-1}}{\mathcal{J}}\left(\frac{A_{j}^{h}}{\calj} \frac{\partial}{\partial a^{h}}\left(\frac{\delta F}{\delta\pi_{j}}\right)\right)\right]
 \mathlarger{\Bigg\}}
 \,.
 \label{eq:FG-mid}
\end{align}
Using the identity of \eqref{IDen} with $z^\ell$ set to $\pi^\ell/\rho_0$,  
\begin{equation}
A_{\ell}^{u}\frac{\partial}{\partial a^{u}}\left[\frac{A_{i}^{k}}{\calj} \frac{\partial}{\partial a^{k}}\left(\frac{\pi^{\ell}}{\rho_{0}}\right)\right]= A_{i}^{k}\frac{\partial}{\partial a^{k}}\left[\frac{A_{\ell}^{u}}{\calj}\frac{\partial}{\partial a^{u}}\left(\frac{\pi^{\ell}}{\rho_{0}}\right)\right]\,, 
\end{equation}
and integrating by parts,   (\ref{eq:FG-mid}) becomes
\begin{align}
[F,G]^D   & =\int\!d^{3}a\,\, \mathlarger{\Bigg\{}
\frac{\delta F}{\delta q^{i}}\left.\mathbb{P}_{\rho_0\perp}\frac{\delta G}{\delta\boldsymbol{\pi}}\right|^{i}-\frac{\delta G}{\delta q^{i}}\left.\mathbb{P}_{\rho_0\perp}\frac{\delta F}{\delta\boldsymbol{\pi}}\right|^{i}
\nonumber \\
 & +\rho_{0}\frac{A_{i}^{k}}{\calj} \frac{\partial}{\partial a^{k}}\left(\frac{\pi^{\ell}}{\rho_{0}}\right)\,\left(\left. 
 \mathbb{P}_{\rho_0}\frac{\delta F}{\delta\boldsymbol{\pi}}\right|^{i}\left.\mathbb{P}_{\rho_0\perp}\frac{\delta G}{\delta\boldsymbol{\pi}}\right|_{\ell} 
 - \left.\mathbb{P}_{\rho_0}\frac{\delta G}{\delta\boldsymbol{\pi}}\right|^{i}\left.\mathbb{P}_{\rho_0\perp}\frac{\delta F}{\delta\boldsymbol{\pi}}\right|_{\ell}\right)
 \nonumber \\
 & -\rho_{0} \frac{A_{\ell}^{u}}{\calj}\frac{\partial}{\partial a^{u}}\left(\frac{\pi^{\ell}}{\rho_{0}}\right)\,\left(\left. 
 \mathbb{P}_{\rho_0}\frac{\delta F}{\delta\boldsymbol{\pi}}\right|^{i}\left.\mathbb{P}_{\rho_0\perp}\frac{\delta G}{\delta\boldsymbol{\pi}}\right|_{i}
 -\left.\mathbb{P}_{\rho_0}\frac{\delta G}{\delta\boldsymbol{\pi}}\right|^{i}\left.\mathbb{P}_{\rho_0\perp}\frac{\delta F}{\delta\boldsymbol{\pi}}\right|_{i}\right)
 \mathlarger{\Bigg\}}\,,
 \label{eq:FG-final}
\end{align}
where we used
\begin{equation}
A_{i}^{k}\frac{\partial}{\partial a^{k}}\mathbb{P}_{\rho_0}\mathbf{z}\,\big|^{i}=A_{i}^{k}\frac{\partial}{\partial a^{k}} \left(\mathbb{P}_{\rho_0}\right)^i_j z^j= 0\,,\qquad\mathrm{for\ all}\  \ \bfz\,,
\label{Ldiv}
\end{equation}
which follows from the definitions \eqref{eq:P0perp}, viz.
\bq
\frac{A_{i}^{k}}{\calj}\frac{\partial}{\partial a^{k}}\left(\mathbb{P}_{\rho_0\perp}\right)^i_j z^{j}=
\frac{A_{i}^{k}}{\calj}\frac{\partial z^{i}}{\partial a^{k}} \,,
\label{divPperp}
\eq
 and \eqref{P}.   Also, upon inserting $\mathbb{P}_{\rho_0\perp}= I - \mathbb{P}_{\rho_0}$ in the last line of \eqref{eq:FG-final}, symmetry implies we can drop the $\mathbb{P}_{\rho_0\perp}$.
Finally, upon substituting (\ref{eq:FG-final}) into (\ref{eq:dbkt1}), we obtain  
\begin{align}
\left\{ F,G\right\} _{*} & =\int\!d^{3}a\,\, \mathlarger{\Bigg\{}
\frac{\delta F}{\delta q^{i}}\left.\mathbb{P}_{\rho_0}\frac{\delta G}{\delta\boldsymbol{\pi}}\right|^{i}-\frac{\delta G}{\delta q^{i}}\left.\mathbb{P}_{\rho_0}\frac{\delta F}{\delta\boldsymbol{\pi}}\right|^{i}
\nonumber \\
 & - \rho_{0}\frac{A_{i}^{k}}{\calj} \frac{\partial}{\partial a^{k}}\left(\frac{\pi^{\ell}}{\rho_{0}}\right)\,\left(\left. 
 \mathbb{P}_{\rho_0}\frac{\delta F}{\delta\boldsymbol{\pi}}\right|^{i}\left.\mathbb{P}_{\rho_0\perp}\frac{\delta G}{\delta\boldsymbol{\pi}}\right|_{\ell} 
 - \left.\mathbb{P}_{\rho_0}\frac{\delta G}{\delta\boldsymbol{\pi}}\right|^{i}\left.\mathbb{P}_{\rho_0\perp}\frac{\delta F}{\delta\boldsymbol{\pi}}\right|_{\ell}\right)
 \nonumber \\
 & +\rho_{0} \frac{A_{\ell}^{k}}{\calj}\frac{\partial}{\partial a^{k}}\left(\frac{\pi^{\ell}}{\rho_{0}}\right)\,\left(\left. 
 \mathbb{P}_{\rho_0}\frac{\delta F}{\delta\boldsymbol{\pi}}\right|^{i} 
 \frac{\delta G}{\delta{\pi^i}}
 -\left.\mathbb{P}_{\rho_0}\frac{\delta G}{\delta\boldsymbol{\pi}}\right|^{i}
 \frac{\delta F}{\delta \pi^i} \right)
 \mathlarger{\Bigg\}}\,.
 \label{eq:dbkt2}
\end{align}
Once more inserting $\mathbb{P}_{\rho_0\perp}= I- \mathbb{P}_{\rho_0}$,  rearranging,  and reindexing gives
\begin{align}
\left\{ F,G\right\} _{*} & =- \int\!d^{3}a\,\rho_0 \, \mathlarger{\Bigg\{}
\frac1{\rho_0}\frac{\delta G}{\delta q^{i}}\left.\mathbb{P}_{\rho_0}\frac{\delta F}{\delta\boldsymbol{\pi}}\right|^{i}
- \frac1{\rho_0} \frac{\delta F}{\delta q^{i}}\left.\mathbb{P}_{\rho_0}\frac{\delta G}{\delta\boldsymbol{\pi}}\right|^{i} 
+   \cala_{mn}  
 \left.\mathbb{P}_{\rho_0\perp} \frac{\delta F}{\delta\boldsymbol{\pi}}\right|^m
 \left.\mathbb{P}_{\rho_0\perp}\frac{\delta G}{\delta\boldsymbol{\pi}}\right|^{n} 
\nonumber \\
 & 
\hspace{ 3cm}  +   \calt_{mn}\,
 \left(
 \frac{\delta F}{\delta{\pi_m}}  \left.\mathbb{P}_{\rho_0\perp}\frac{\delta G}{\delta\boldsymbol{\pi}}\right|^{n} 
 -   \frac{\delta G}{\delta \pi_m}
 \left.\mathbb{P}_{\rho_0\perp}\frac{\delta F}{\delta\boldsymbol{\pi}}\right|^{n}
 \right)
 \mathlarger{\Bigg\}}\,, 
 \label{eq:dbkt-2} 
\end{align}
where
\bq
\cala_{nm}:=    \eta_{\ell m} D^\ell_n-   \eta_{\ell n} D^\ell_m
  \qquad 
\mathrm{and}
\qquad
\calt_{mn}:=  \eta_{\ell n}  D^\ell_m + \eta_{mn}   D^2 \,,
\label{CB}
\eq
with
\bq
D_m^{\ell}= \frac{A_{m}^{k}}{\calj} \frac{\partial}{\partial a^{k}} \left(\frac{\pi^{\ell}}{\rho_{0}}\right)\,.
\eq
Note the trace $D^\ell_\ell= D^2$, which we will eventually set to zero. Equation \eqref{eq:dbkt-2} gives  the Dirac bracket for the incompressibility  holonomic constraint.  This bracket  with the Hamiltonian
\bq
H=\int \! d^3a \, \frac{|\bfpi|^2}{2\rho_0} =\int \! d^3a \,   \eta^{mn} \frac{\pi_m \pi_n}{2\rho_0}   \,,
\label{ham}
\eq
produces  dynamics that fixes $\calj$ and thus enforces  incompressibility provided the constraint $D^2=0$ is used as an initial condition. 
For MHD we add to $H$ the following:
\bq
H_B=\int \! d^3a \,  \eta_{mn}  \frac{B_0^j B_0^k }{2\calj}\,\frac{\p q^m}{\p a^j} \frac{\p q^n}{\p a^k}  \,.
\eq
We note, any Hamiltonian that is consistent with \eqref{constraints-1-1} can be used to define a constrained flow. 

Proceeding to the equations of motion,  we first calculate $\dot{q}^i$, 
\bal
\dot{q}^i&= \{q^i,H\}_{*} 
= \left(\mathbb{P}_{\rho_0}\right)^i_j \frac{\delta H}{\delta \pi_j}
=   \frac{\delta H}{\delta \pi_i}  - \frac{\eta^{i\ell}}{\rho_{0}} A_{\ell}^{u}  \frac{\partial}{\partial a^{u}}\left[\frac{\Delta_{\rho_{0}}^{-1}}{\mathcal{J}}\left(\frac{A_{j}^{h}}{\calj}\frac{\partial}{\partial a^{h}} \frac{\delta H}{\delta \pi_j}\right)\right]
\nonumber \\
& =  \frac{\pi^i}{\rho_0}  - \frac{\eta^{i\ell}}{\rho_{0}} A_{\ell}^{u}  \frac{\partial}{\partial a^{u}}\left[\frac{\Delta_{\rho_{0}}^{-1}}{\mathcal{J}}\left(\frac{A_{j}^{h}}{\calj}\frac{\partial}{\partial a^{h}} \frac{\pi^j}{\rho_0}\right)\right]\,.
\label{LDeomQ}
\eal
The equation for $\dot{\pi}_i$ is more involved.  Using the adjoint property of \eqref{PperpSA}, which is valid for both $\mathbb{P}_{\rho_0\perp}$ and 
$\mathbb{P}_{\rho_0}$,  we obtain 
\bal
 \dot{\pi_i}&= \{\pi_i,H\}_{*} =- \rho_0 \left( \mathbb{P}_{\rho_0}\right)^j_i \frac1{\rho_0} 
 \frac{\delta H}{\de q^j}  
-   \rho_{0}\,  \left(\mathbb{P}_{\rho_0\perp}\right)^m_i 
  \left(
\cala_{mn} 
 \left(\mathbb{P}_{\rho_0\perp} \right)^n_k
\frac{\delta H}{\delta \pi_k}
 \right)
 \nonumber\\
 &\hspace{2cm} +\rho_{0}\,  \left(\mathbb{P}_{\rho_0\perp} \right)^n_i 
  \left(
 \calt_{mn}
\frac{\delta H}{\delta \pi_m}
 \right)
-  \rho_{0}\,   \calt_{in} \left(\mathbb{P}_{\rho_0\perp} \right)^n_k 
\frac{\delta H}{\delta \pi_k}
 \nonumber\\
 & =  
   - \rho_{0}\,  \left(\mathbb{P}_{\rho_0\perp}\right)^m_i 
  \left(
\cala_{mn} 
 \left(\mathbb{P}_{\rho_0\perp} \right)^n_k
\frac{\pi^k}{\rho_0}
 \right)
+  \rho_{0}\,  \left(\mathbb{P}_{\rho_0\perp} \right)^n_i 
  \left(
 \calt_{mn}
\frac{\pi^m}{\rho_0}
 \right)
-  \rho_{0}\,   \calt_{in} \left(\mathbb{P}_{\rho_0\perp} \right)^n_k 
\frac{\pi^k}{\rho_0}\,,
\label{LDeomP}
\eal
which upon substitution of the definitions of $\mathbb{P}_{\rho_0}$, $\cala_{mn}$, and $\calt_{mn}$ of \eqref{eq:P0perp} and \eqref{CB}  yields a complicated nonlinear equation.

Equations  \eqref{LDeomQ} and \eqref{LDeomP} are infinite-dimensional  versions of the finite-dimensional systems of  \eqref{eomq} and \eqref{eomp} considered in  Section \ref{sssec:HCD}. There,  equations  \eqref{eomq} and \eqref{eomp} were reduced to \eqref{eomqr} and \eqref{eompr} upon enforcing the holomomic constraint by requiring that initially $D^2=0$.  Similarly we can enforce the vanishing of $D^2$ of  \eqref{constraints-1-1}, which is compatible with  the  Hamiltonian \eqref{ham}.   Instead of  addressing this evaluation now, we find the  meaning of various terms  is much more transparent when written in terms of  Eulerian variables, which we do in Section \ref{ssec:EDcon}. We  then return to these Lagrangian equations in Section \ref{ssec:comparison} and make comparisons.   Nevertheless, the solution of equations \eqref{LDeomQ} and  \eqref{LDeomP}, $\bfq(\bfa,t)$,  with  the initial conditions $D^1=-\ln \rho_0$ and $D^2=0$, is  a volume preserving transformation at any time $t$.

\subsection{Eulerian-Dirac constraint theory}
\label{ssec:EDcon}

Because we chose the form of constraints $D^{1,2}$ of \eqref{D1} and \eqref{constraints-1-1}  to be Eulerianizable,  it follows that we can transform easily the results of Section \ref{sssec:LDholo} into Eulerian form.  This we do in Section \ref{sssec:LDEP}. Alternatively, we can proceed as in  \citet{turski99,turski01,pjmTC09,pjmCGBT13,pjmLB09}, starting from the Eulerian noncanonical theory of Section \ref{ssec:LtoE} and directly construct a Dirac bracket with Eulerian  constraints.  This is a valid procedure because Dirac's construction works for noncanonical Poisson brackets, as shown, e.g.,  in \citet{pjmLB09}, but it does not readily allow for advected density.  This direct method with uniform density  is reviewed in Section \ref{sssec:EDD}, where it is  contrasted with the results of Section \ref{sssec:LDEP}.

\subsubsection{Lagrangian-Dirac constraint theory in the Eulerian picture}
\label{sssec:LDEP}

In  a  manner similar to that used to obtain \eqref{deFq} and \eqref{deFpi},  we find  the  functional derivatives transform as
\begin{equation}
\frac{\delta F}{\delta\pi_{i}}=\frac{1}{\rho}\frac{\delta \bar{F}}{\delta \varv_{i}},\quad\frac{1}{\rho_{0}}\frac{\delta F}{\delta q^{i}}=\frac{\partial}{\partial x^{i}}\frac{\delta \bar{F}}{\delta\rho}-\frac{1}{\rho}\frac{\delta \bar{F}}{\delta s}\frac{\partial s}{\partial x^{i}}-\frac{1}{\rho}\frac{\delta \bar{F}}{\delta \varv_{\ell}}\frac{\partial \varv_{\ell}}{\partial x^{i}} \,,
\label{chain}
\end{equation}
where the expressions on the left of each equality are clearly Lagrangian variable quantities, while on the right they are Eulerian quantities represented  in terms of  Lagrangian variables.   
Substituting these expressions into (\ref{eq:dbkt}) and dropping the bar on $F$ and $G$ gives the following bracket in terms of the Eulerian variables:
\begin{align}
\left\{ F,G\right\} _{*} & =-\int\!d^{3}x\,\, \mathlarger{\Bigg\{}
\left(\left.
\mathcal{P}_{\rho}\frac{\delta F}{\delta\mathbf{v}}\right|^{i} \frac{\partial}{\partial x^{i}}\frac{\delta G}{\delta\rho} 
- \left.\mathcal{P}_{\rho}\frac{\delta G}{\delta\mathbf{v}}\right|^{i}\frac{\partial}{\partial x^{i}}\frac{\delta F}{\delta\rho}\right)
\nonumber \\
 & +\frac{1}{\rho}\frac{\partial s}{\partial x^{i}}
 \left(\frac{\delta F}{\delta s}\left.\mathcal{P}_{\rho}\frac{\delta G}{\delta\mathbf{v}}\right|^{i} 
 - \frac{\delta G}{\delta s}\left.\mathcal{P}_{\rho}\frac{\delta F}{\delta\mathbf{v}}\right|^{i}\right)
 \nonumber \\
 & +\frac{1}{\rho}\frac{\partial \varv^{\ell}}{\partial x^{i}}\left(\left.\mathcal{P}_{\rho}\frac{\delta F}{\delta\mathbf{v}}\right|_{\ell}\left.\mathcal{P}_{\rho}\frac{\delta G}{\delta\mathbf{v}}\right|^{i} 
 -\left.\mathcal{P}_{\rho}\frac{\delta G}{\delta\mathbf{v}}\right|_{\ell}\left.\mathcal{P}_{\rho}\frac{\delta F}{\delta\mathbf{v}}\right|^{i}\right) 
 \nonumber \\
 & +\frac{1}{\rho}\frac{\partial \varv^{\ell}}{\partial x^{\ell}}\,\left(\left.\mathcal{P}_{\rho}\frac{\delta F}{\delta\mathbf{v}}\right|^{i}\frac{\delta G}{\delta \varv_{i}}
 -\left.\mathcal{P}_{\rho}\frac{\delta G}{\delta\mathbf{v}}\right|^{i}\frac{\delta F}{\delta \varv_{i}}\right)
  \mathlarger{\Bigg\}}\,,
  \label{eq:dbkt-3}
\end{align}
where we used the relations \eqref{vol} and \eqref{ELgrad} 
and we introduced the Eulerian projection operator
\bq
 \left.\mathcal{P}_{\rho} \frac{\de F}{\de \bfv}\right|^i =  (\mathcal{P}_{\rho})^i_j\frac{\de F}{\de \varv_j}
 = \left. \rho\mathbb{P}_{\rho_0}\frac{\delta F}{\delta\boldsymbol{\pi}}\right|^i
 \qquad
 \mathrm{and}
 \qquad
 \left.\mathcal{P}_{\rho} \frac{\de F}{\de \bfv}\right|_i =\eta_{ij}\left.\mathcal{P}_{\rho} \frac{\de F}{\de \bfv}\right|^j \,,
\eq
with
\bq
\left(\mathcal{P}_{\rho}\right)^i_j z^j=\de^i_j - \eta^{ik} \frac{\p}{\p x^k} \left[\De_\rho^{-1} \frac{\p }{\p x^j} 
\left(\frac{z^j}{\rho}\right)\right]\,,
\eq
which is easily seen to satisfy $(\mathcal{P}_{\rho})^i_j (\mathcal{P}_{\rho})^j_k=(\mathcal{P}_{\rho})^i_k$.
Observe, like its Lagrangian counterpart, $\mathcal{P}_{\rho}$ is formally self-adjoint; however, this time we found  it convenient to  define the projection in  such a way that the self-adjointness is with respect to a different weighted inner product, viz. 
\bq
\int \frac{d^3x}{\rho} \, w_i \, (\mathcal{P}_{\rho })^i_j \, z^j = \int \frac{d^3x}{\rho} \, z_i \, (\mathcal{P}_{\rho})^i_j \, w^j \,.
\eq

In terms of usual cartesian vector notation
\begin{equation}
\mathcal{P}_{\rho}\frac{\delta G}{\delta\mathbf{v}}=\frac{\delta G}{\delta\mathbf{v}}-\nabla\Delta_{\rho}^{-1}\nabla\cdot\left(\frac{1}{\rho}\frac{\delta G}{\delta\mathbf{v}}\right) \,.
\label{eq:Peul}
\end{equation}
Upon writing   $\mathcal{P}_{\rho}=I-\mathcal{P}_{\rho \perp}$ and  decomposing an arbitrary  vector field as 
\[
\mathbf{z}=-\nabla\Phi+\rho\nabla\times\mathbf{A},
\]
this projection operator yields the component $\mathcal{P}_{\rho}\mathbf{z}=\rho\nabla\times\mathbf{A}$. 
 {Therefore,  if  $\nabla\rho\times\mathbf{A}=0$,  then this  operator  projects into the space of incompressible vector fields.}  For convenience  we introduce the associated projector 
\bq
\mathbb{P}_\rho \bfv :=  \bfv -\frac1{\rho} \nabla\Delta_{\rho}^{-1}\nabla\cdot \bfv 
= \frac1{\rho} {\calp_{\rho}}(\rho \bfv)  \,, 
\label{Pmap}
\eq
which has the desirable property  
\bq
\nabla\cdot (\mathbb{P}_\rho  \bfv)=0   \quad \forall\  \mathbf{v}
\qquad\mathrm{compared\ \ to}\qquad 
 \nabla\cdot \left(\frac1{\rho} {\calp_{\rho}} \bfw\right)=0 \quad \forall\  \mathbf{w}\,.
\label{divP}
\eq
Upon writing   $\mathbb{P}_{\rho}=I-\mathbb{P}_{\rho \perp}$ and  decomposing an arbitrary  vector field $\bfv$ as 
\[
\mathbf{v}=-\frac1{\rho}\, \nabla\Phi+ \nabla\times\mathbf{A},
\]
this projection operator yields the component $\mathbb{P}_{\rho}\mathbf{v}= \nabla\times\mathbf{A}$, while $\mathbb{P}_{\rho \perp}\bfv = \nabla \Phi/\rho$.   
Note, $\mathbb{P}_\rho$ is the Eulerianization of 
$\mathbb{P}_{\rho_0}$ and it is not difficult to write \eqref{eq:dbkt-4} in terms of this quantity.

Upon  adopting this  usual vector notation, the bracket (\ref{eq:dbkt-3})
can also be written as 
\begin{eqnarray}
\left\{ F,G\right\} _{*} & = & -\int\,d^{3}x\,\left[\nabla\frac{\delta G}{\delta\rho}\cdot\mathcal{P}_{\rho} \frac{\delta F}{\delta\mathbf{v}}-\nabla\frac{\delta F}{\delta\rho}\cdot\mathcal{P}_{\rho} \frac{\delta G}{\delta\mathbf{v}}\right.
\nonumber \\
 &  & +\frac{\nabla s}{\rho}\cdot\left(\frac{\delta F}{\delta s}\mathcal{P}_{\rho} \frac{\delta G}{\delta\mathbf{v}}-\frac{\delta G}{\delta s}\mathcal{P}_{\rho} \frac{\delta F}{\delta\mathbf{v}}\right)
 \nonumber \\
 &  & +\frac{\nabla\times\mathbf{v}}{\rho}\cdot\left(\mathcal{P}_{\rho} \frac{\delta G}{\delta\mathbf{v}}\times\mathcal{P}_{\rho} \frac{\delta F}{\delta\mathbf{v}}\right)
 \nonumber \\
 &  & +\left.\frac{\nabla\cdot\mathbf{v}}{\rho}\left(\frac{\delta F}{\delta\mathbf{v}}\cdot 
 \mathcal{P}_{\rho} \frac{\delta G}{\delta\mathbf{v}} 
 -\frac{\delta G}{\delta\mathbf{v}}\cdot\mathcal{P}_{\rho} \frac{\delta F}{\delta\mathbf{v}}\right)\right]\,.
 \label{eq:dbkt-4}
\end{eqnarray}
For MHD there is a magnetic field contribution to \eqref{chain} and following the steps that lead  to  \eqref{eq:dbkt-4} we obtain 
\bal
 \{F,G\}_{*B} & =  -\int \!d^3 x \, \Bigg[
 \bfB\cdot
    \left(
    \frac{1}{\rho} \mathcal{P}_{\rho}\frac{\delta F}{\delta \bfv} \cdot\nabla \frac{\delta G}{\delta \bfB}
-    \frac{1}{\rho} \mathcal{P}_{\rho}\frac{\delta G}{\delta \bfv} \cdot\nabla \frac{\delta F}{\delta \bfB}
\right)
\nonumber\\
&\hspace{1cm}
+\bfB\cdot
    \left(
 \nabla \left(\frac{1}{\rho}\mathcal{P}_{\rho} \frac{\delta F}{\delta \bfv} \right ) \cdot \frac{\delta G}{\delta \bfB}
-    \nabla \left(\frac{1}{\rho}\mathcal{P}_{\rho} \frac{\delta G}{\delta \bfv} \right ) \cdot \frac{\delta F}{\delta \bfB}
\right)
            \Bigg]\,.
 \label{PvBbkt}
\eal 
With the exception of the last term of \eqref{eq:dbkt-4} proportional to $\nabla\cdot \bfv$ and the presence of  the  Eulerian projection operator $\calp_\rho$,  \eqref{eq:dbkt-4} added to \eqref{PvBbkt} is  identical to the noncanonical Poisson bracket for the ideal fluid and MHD as given in  \citet{pjmG80}. 
 By construction, we know that \eqref{eq:dbkt-4} satisfies the Jacobi identity --  this follows because it was obtained by Eulerianizing  the  canonical Dirac bracket in terms of Lagrangian variables. 
Guessing the bracket and proving Jacobi for \eqref{eq:dbkt-4} directly would be a difficult chore, giving credence to  the path we have followed in obtaining it. 

To summarize, the bracket of \eqref{eq:dbkt-4} together with the Hamiltonian
\bq
H=\frac12\int d^3x\, \rho \,|\bfv|^2\,,
\label{HamAgain}
\eq
the Eulerian counterpart of \eqref{ham},  generates dynamics that can preserve the constraint $\nabla\cdot\bfv=0$.  If we add $H_B=\int d^3x\,  |\bfB|^2/2$ to \eqref{HamAgain} and add \eqref{PvBbkt} to  \eqref{eq:dbkt-4}, then we obtain incompressible MHD.   The fluid case  is the Eulerian counterpart of the volume preserving geodesic flow, described  originally by   Lagrange in Lagrange variables. Upon performing a series of straightforward manipulations,  we obtain the following equations of motion for the flow: 
\bal
\frac{\p \rho}{\p t}&= \left\{ \rho,H\right\} _{*} = -\nabla    \cdot\mathcal{P}_{\rho}\frac{\delta H}{\delta\mathbf{v}} = -\nabla \rho  \cdot  \mathbb{P}_\rho \bfv
\,,
\label{Pden}\\
\frac{\p s}{\p t}&= \left\{s,H\right\} _{*} = -\frac{\nabla s}{\rho}\cdot\mathcal{P}_{\rho}\frac{\delta H}{\delta\mathbf{v}}
= - \nabla s  \cdot  \mathbb{P}_\rho \bfv\,,
\label{Pent}\\
\frac{\p \bfv}{\p t}&= \left\{ \bfv,H\right\} _{*} =  -\frac1{\rho} \calp_{\rho} \left( \rho \nabla \frac{\delta H}{\delta \rho}\right)
 + \frac1{\rho}  \calp_{\rho}\left( {\nabla s}\frac{\delta H}{\delta s}\right) 
-\frac1{\rho}  \mathcal{P}_{\rho} 
\left((\nabla\times\mathbf{v}) \times  \mathcal{P}_{\rho} \frac{\delta H}{\delta\mathbf{v}} 
\right) 
\nonumber\\
& - \frac{\nabla\cdot \bfv}{\rho}  \mathcal{P}_{\rho} \frac{\delta H}{\delta\mathbf{v}}
+ \frac1{\rho}  \mathcal{P}_{\rho} \left( {\nabla\cdot \bfv} \,  \frac{\delta H}{\delta\mathbf{v}}
\right)
\nonumber\\
&= - \P_\rho \nabla \frac{|\bfv|^2}{2} - \P_\rho \left( (\nabla\times \bfv)\times \P_\rho \bfv\right) -  (\nabla\cdot \bfv)\,   \P_\rho \bfv
+  \,   \P_\rho \left(\bfv\, \nabla\cdot \bfv\right)\,.
\label{pvt}
\eal
If we include $H_B$ we  obtain  additional terms to \eqref{pvt} generated by \eqref{PvBbkt} for the projected $\bfJ\times\bfB$ force. 
Observe,  equation \eqref{pvt} is not yet evaluated on the constraint $D^2=0$, which in Eulerian variables is $\nabla\cdot\bfv=0$. 
As noted at the end of  Section \ref{ssec:LDconTh}, we turn to this task in Section \ref{ssec:comparison}.

\subsubsection{Eulerian-Dirac constraint theory direct with uniform density}
\label{sssec:EDD}

For completness we recall  the simpler case  where the Eulerian density $\rho$ is  uniformly  constant, which  without loss of generality can be scaled to unity.  This case was  considered  in  \citet{turski99,turski01,pjmCT12,pjmCGBT13} (although a trick of using entropy as density was employed in \citet{pjmCGBT13} to treat density advection).  In these works the Dirac constraints were chosen to be the pointwise  Eulerian quantities 
\bq
\cald^1=\rho  \qquad \mathrm{and} \qquad  \cald^2=\nabla\cdot \bfv 
\,,
\label{Edirac}
\eq
and the Dirac procedure was effected on the purely Eulerian level. 
This led to the projector
\bq
\P:=\P_{\rho=1}=1 -\nabla \Delta^{-1}\nabla\,\cdot  \ \,, 
\label{eq:P}
\eq 
where  $\De=\De_{\rho=1}$, and  the following Dirac bracket: 
\begin{eqnarray}
 \{F,G\}_*&=&- \int d^3x\,  \Bigg[ 
\frac{\nabla s}{ \rho} \cdot \left( \frac{\de F}{\de s} \P\frac{\de G}{\de \bf v} 
- \frac{\de G}{\de s} \P\frac{\de F}{\de \bf v} \right)
\nonumber\\
&& \hspace{1cm} -  \frac{\nabla\times {\bf v}}{\rho}\cdot \left( \P\frac{\de F}{\de \bf v}\times \P\frac{\de G}{\de \bf v}\right) \Bigg]\,.
 \label{eqn:PBD}
\end{eqnarray}
Incompressible MHD with constant density is generated by adding the following to  \eqref{eqn:PBD}
\begin{eqnarray}
 \{F,G\}_{*B}&=&- \int d^3x\,  \Bigg[\frac{\bf B}{\rho} \cdot \left(\P\frac{\de F}{\de \bf v} \cdot \nabla \frac{\de G}{\de \bf B} 
-\P\frac{\de G}{\de \bf v} \cdot \nabla \frac{\de F}{\de \bf B} \right)
 \nonumber \\
  && \hspace{1cm} + {\bf B}\cdot \left(\nabla \left(\frac1{\rho}\P\frac{\de F}{\de \bf v}\right)  \cdot \frac{\de G}{\de \bf B}-  \nabla \left(\frac1{\rho} \P\frac{\de G}{\de \bf v}\right) \cdot \frac{\de F}{\de \bf B}\right)\Bigg]\,,
 \label{eqn:BPBD}
\end{eqnarray}
and adding $|\bfB|^2/2$ to the integrand of \eqref{HamAgain}. 

The bracket of \eqref{eqn:PBD} differs from that of \eqref{eq:dbkt-4} in two ways: the projector $\mathcal{P}_{\rho}$ is replaced by the simpler projector $\P$  and it is missing the term proportional to $\nabla\cdot \bfv$.  Given that $\nabla\cdot \bfv$ cannot be set to zero until after the equations of motion are obtained, this term gives rise to a significant differences between  the constant and nonconstant density Poisson brackets and incompressible dynamics.

\subsection{Comparison of the Eulerian-Dirac and Lagrangian-Dirac constrained theories}
\label{ssec:comparison}

 Let us now discuss equations \eqref{Pden}, \eqref{Pent} and \eqref{pvt}.  Given that $\nabla\cdot \mathbb{P}_{\rho}\mathbf{v}=0$ (cf.\ \eqref{divP})
 it is clear that the density and entropy are advected by the incompressible velocity field $\mathbb{P}_{\rho}\mathbf{v}$, as expected.   However, the meaning of  \eqref{pvt} remains to be clarified.  To this end we take  the divergence of  \eqref{pvt}  and again use  \eqref{divP} to obtain 
\bq
\frac{\p(\nabla\cdot \bfv)}{\p t} = -\nabla\cdot\left(\nabla\cdot \bfv \,  \P_\rho \bfv\right)
=  -\left(\P_\rho \bfv\right) \cdot  \!\nabla\,  (\nabla \cdot    \bfv)\,.
\label{divvt}
\eq
Thus $\nabla\cdot\bfv$ itself is advected by an incompressible velocity field.  As with any advection equation, if initially  
$\nabla\cdot\bfv =  0$ , it will remain uniformly zero.   After setting $\nabla\cdot \bfv =0$ in   equation \eqref{pvt} it collapses down to 
\bq
\frac{\p \bfv}{\p t}= - \P_\rho \left(\bfv\cdot \nabla \bfv\right)\,;
\label{Eeom}
\eq
this is the  anticipated equation of motion, the momentum equation of \eqref{momentum} with the insertion of the pressure given by \eqref{presrho}.

Given the  discussion of Lagrangian vs.\ Eulerian constants of motion  of Section \eqref{ssec:CoM},  that $\nabla\cdot \bfv$ is advected rather than pointwise conserved is to be expected.  Our development began with the constraints $D^{1,2}$ of \eqref{D1} and \eqref{constraints-1-1} both of which are pointwise conserved by the Dirac procedure, i.e.\ $\dot\cald_L\equiv 0$.  This means their corresponding  fluxes are identically zero, i.e., in  \eqref{Lcon}  we have  $\mathbf{\Ga}_{\cald_L}\equiv 0$ for each.     Thus the flux component  $\mathbf{ \bar\Ga}_{\cald_E}$ of \eqref{ELcon} vanishes and the Eulerian flux for both $D^1$ and $D^2$ have the form  $\bfv {\cald_E}$.  Because $D^1$ and $D^2$ Eulerianize to $-\ln(\rho)$ and $\nabla\cdot\bfv$, respectively, we expected equations of the from of \eqref{divvt} for both.  We will see in Section \ref{ssec:AoI} that the equation for $D^1$  in fact follows also because  the constraints are  Casimir invariants.

Let us return to \eqref{LDeomP} and compare with the results of Section \ref{sssec:HCD}.  Because the incompressibility condition is an holonomic  constraint and Section \ref{sssec:HCD} concerns holonomic constraints for the uncoupled $N$-body problem, both results are geodesic flows.   In fact, one can think of the fluid case as a continuum version of that of Section \ref{sssec:HCD} with an infinity of holonomic constraints-- thus  we expect similarities between these results.  However, because the incompressibility constraints are pointwise constraints, the comparison is not as straightforward as it would be for global constraints of the fluid. 

To make the comparison we first observe that the  term $\cala_{mn}$ of \eqref{eq:dbkt-2} must correspond to the term $\overset\leftrightarrow{\mathbb{A}}_{ij}$ of \eqref{TA}, since their origin follows an analogous path in the derivation, both are antisymmetric, and both project  from both the left and the right.   
The analog of \eqref{finP} according to \eqref{eq:P0perp} is 
\bq
\left(\mathbb{P}_{\rho_0\perp}\right)^{i}_j\, \frac{\pi^j}{\rho_0}= 
\frac{\eta^{i\ell}}{\rho_{0}} A_{\ell}^{u}  \frac{\partial}{\partial a^{u}}\left[\frac{\Delta_{\rho_{0}}^{-1}}{\mathcal{J}}\left(\frac{A_{j}^{h}}{\calj}\frac{\partial}{\partial a^{h}}\, \frac{\pi^j}{\rho_0}\right)\right]
= \frac{\eta^{i\ell}}{\rho_{0}} A_{\ell}^{u}  \frac{\partial}{\partial a^{u}}\left[\frac{\Delta_{\rho_{0}}^{-1}}{\mathcal{J}}\left(D^2\right)\right]
\equiv 0 \,,
\label{infinP}
\eq
when evaluated on $D^2=0$.  Unlike \eqref{finP} a sum, which would here be an integral over $d^3a$, does not occur because the constraint $D^2$ is a pointwise constraint as opposed to a global constraint.  Also, because the constraints are pointwise,  the $\overset\leftrightarrow{\mathbb{T}}_{ij}$ is analogous to the terms with $\calt_{mn}$ that also have a factor of the projector  $\mathbb{P}_{\rho_0\perp}$, giving the results analogous to \eqref{finT}.
Just as in Section \ref{sssec:HCD}, we obtain $\pi^i=\rho_0\dot{q}^i$ from \eqref{LDeomQ} when evaluated on the constraint $D^2=0$ and only a single term involving the $\calt_{mn}$ contributes to the momentum equation of motion  \eqref{LDeomP}.   {We obtain
\bal
\dot{\pi}_i&=  \rho_0 \eta_{in} \left(\mathbb{P}_{\rho_0\perp}\right)^{n}_r\,
\Big(
\dot{q}^m\, \eta^{rs}  \calt_{ms}
\Big)
\nonumber\\
&= A^u_i\frac{\p }{\p a^u}
\left\{\frac{\Delta_{\rho_{0}}^{-1}}{\mathcal{J}} \left[\frac{A_{\ell}^{h}}{\calj} \frac{\partial}{\partial a^{h}} 
  \left( \dot{q}^m\, 
  \frac{A^k_m}{\calj} \frac{\p \dot{q}^\ell}{\p a^k}
  \right) \right]\right\} 
\nonumber\\
&= A^u_i\frac{\p }{\p a^u}
\left\{\frac{\Delta_{\rho_{0}}^{-1}}{\mathcal{J}} \left[\frac{A_{\ell}^{h}}{\calj} \frac{\partial}{\partial a^{h}} 
  \left(
  \frac{A^k_m}{\calj} \frac{\p (\dot{q}^m\dot{q}^\ell)}{\p a^k}
  \right) \right]\right\} \,,
  \label{pidot}
\eal
where  the second equality follows upon substitution of 
\[
\calt_{ms} \rightarrow \eta_{\ell s} \frac{A^k_m}{\calj} \frac{\p \dot{q}^\ell}{\p a^k}\,,
\quad \mathrm{for} \quad D^2=0\,,
\]
which follows from \eqref{CB}, while the third follows again from $D^2=0$ according to  \eqref{constraints-1-1}.  Thus,
\bal
\ddot{q}^i&= \eta^{i\ell} \frac{A^u_\ell}{\rho_0} \frac{\p }{\p a^u}
\left\{\frac{\Delta_{\rho_{0}}^{-1}}{\mathcal{J}} 
\left[\frac{A_{j}^{h}}{\calj} \frac{\partial}{\partial a^{h}} \left(\frac{A^f_k}{\calj}
 \frac{\p (\dot{q}^j\, \dot{q}^k) }{\p a^f} 
  \right) \right]\right\}
  \nonumber\\
  &
   = \left(\mathbb{P}_{\rho_0\perp} \right)^{i}_j\, 
    \left(\frac{A^f_k}{\calj}
 \frac{\p (\dot{q}^j\, \dot{q}^k) }{\p a^f} 
  \right)
=: -\, \widehat{\Ga}^{\,i}_{jk}(\dot{q}^j, \dot{q}^k)\,,
   \label{infPCS}
  \eal
where in \eqref{infPCS} we have defined $ \widehat{\Ga}^{\,i}_{jk}(\dot{q}^j, \dot{q}^k)$,  the normal force operator for geodesic flow, analogous to that of \eqref{finalD}.} 

{As was the case for the   $ \widehat{\boldsymbol{\Ga}}_{i,jk}$ of  \eqref{PCS}, $ \widehat{\Ga}^{\,i}_{jk}$ possesses  symmetry: given arbitrary vector fields $\bfV$ and $\bfW$ 
 \bq
  \widehat{\Ga}^{\,i}_{jk}(V^j, W^k):= -\eta^{i\ell} \frac{A^u_\ell}{\rho_0} \frac{\p }{\p a^u}
\left\{\frac{\Delta_{\rho_{0}}^{-1}}{\mathcal{J}} 
\left[\frac{A_{j}^{h}}{\calj} \frac{\partial}{\partial a^{h}} \left(\frac{A^f_k}{\calj}
 \frac{\p (V^j\, W^k) }{\p a^f} 
  \right) \right]\right\}  = \widehat{\Ga}^{\,i}_{jk}(V^k, W^j)\,.
   \label{GA}
  \eq
where the second equality follows from the commutation relation of \eqref{IDen}.}

Equation  \eqref{infPCS} defines geodesic flow on the group of volume preserving diffeomorphisms,  as was the case in Section \ref{sssec:HCD}, it  does so in terms of the original coordinates, i.e., without specifically transforming to normal coordinates on the constraint surfaces which here are infinite-dimensional.

Now we are in position to  close the circle by writing \eqref{infPCS} in Eulerian form.  We will do this for the ideal fluid, but MHD follows similarly.   {As usual the term $\ddot{q}^i$ becomes  the advective derivative $\p \bfv/\p t +\bfv\cdot\nabla\bfv$,  the projector  $\mathbb{P}_{\rho_0\perp}$ becomes $\mathbb{P}_{\rho\perp}$ (using $\De_{\rh_0}^{-1}=\calj \De_{\rh}^{-1}$) when Eulerianized, and  
}
the  $ \widehat{\Ga}^i_{jk}$ term becomes  $\P_{\rho \perp}\left(\nabla\cdot (\bfv\otimes \bfv)\right)$.  Thus  \eqref{infPCS} is precisely the Lagrangian form of \eqref{Eeom}, written as follows:
\bq
\frac{\p \bfv}{\p t}= - \P_\rho \big(\nabla\cdot (\bfv\otimes \bfv)\big)= - \P_\rho \left(\bfv\cdot \nabla \bfv\right)\,.
\eq
Similarly, the Lagrangian version of \eqref{divvt} follows easily from  \eqref{infPCS}.  To see this we operate with the counterpart of taking the Eulerian divergence on the first line of  \eqref{pidot} and make use of   \eqref{divPperp}, 
\bal
\frac{A^h_n}{\calj} \frac{ \p}{\p a^h} \frac{\dot{\pi}^n}{\rho_0} 
&=  \frac{A^h_n}{\calj} \frac{ \p}{\p a^h} 
 \left(\mathbb{P}_{\rho_0\perp}\right)^{n}_r\,
\Big(
\dot{q}^m\, \eta^{rs}  \calt_{ms}
\Big)
=
 \frac{A^h_n}{\calj} \frac{ \p}{\p a^h} 
\Big(
\dot{q}^m\, \eta^{ns}  \calt_{ms}
\Big)
\nonumber\\
&=  \de^n_\ell \frac{A^h_n}{\calj} \frac{ \p}{\p a^h}
 \left(
\dot{q}^m \frac{A^k_m}{\calj} \frac{\p \dot{q}^\ell}{\p a^k} 
\right) \,,
\eal
which in Eulerian variables becomes
\bq
\nabla\cdot\left(\frac{\p \bfv}{\p t} + \bfv\cdot\nabla\bfv \right)
= \nabla\cdot\left(\bfv\cdot\nabla\bfv \right)\qquad \mathrm{or} \qquad 
\nabla\cdot\frac{\p \bfv}{\p t} =\frac{\p\, }{\p t} \nabla\cdot\bfv =0\,.
\eq
In  Lagrangian variables we have the trivial conservation laws
\bq
\dot{\rho_0}= 0 \qquad\mathrm{and} \qquad \dot{s_0}= 0 \,,
\eq
where the corresponding fluxes are identically zero.  However, as is evident from \eqref{Pden} and \eqref{Pent} we obtain nontrivial conservation laws for $\rho$ and $s$ with nonzero fluxes.  Thus we see again, consistent with Section \ref{ssec:CoM}, how Lagrangian and Eulerian conservation laws are not equivalent. 

For the special case where $\rho_0=\calj=1$ one could proceed directly from \eqref{momentumCl}, write it in terms of the Lagrangian variables,  and obtain   \eqref{infPCS}. However, without the constraint theory, one would not immediately see it is Hamiltonian and in fact geodesic flow on an infinite-dimensional manifold.

\subsection{Incompressible algebra of invariants}
\label{ssec:AoI}

In closing this section, we examine the constants of motion for the constrained system.
The Poisson bracket together with the set of functionals that commute with the Hamiltonian, i.e., that satisfy $\{H, I_a\}=0$ for $a = 1, 2,\dots, d$,  constitute the $d$-dimensional  algebra of invariants, a subalgebra of the infinite-dimensional Poisson bracket realization on all functionals. This subalgebra is a Lie algebra realization associated with a symmetry group of the dynamical system, and the Poisson bracket with $\{I_a, \cdot\,  \}$ yields the infinitesimal generators of the symmetries, i.e., the differential operator realization of the algebra.  This was shown for compressible MHD in  \cite{pjm82},   where the associated  Lie algebra realization of the 10 parameter Galilean group on functionals  was  described.  This algebra is homomorphic to usual representations of the Galilean group, with the Casimir invariants being in the center  of the algebra composed of elements that have vanishing Poisson bracket with all other elements. 

A natural question to ask is what happens to this algebra when  incompressibility is enforced by our Dirac constraint procedure.  Obviously the Hamiltonian  is in the subalgebra and  $\{H, \cdot\,  \}_*$  clearly generates time translation, and this will be true for any Hamiltonian, but here we use Hamiltonian of \eqref{HamAgain}. 

Inserting  the momentum
\bq
\bfP=\int \!d^3x \, \rho \bfv
\eq
into \eqref{eq:dbkt-4}  with the Hamiltonian  \eqref{HamAgain} gives
\bq
\{\bfP, H\}_* = 0
\label{PH}
\eq
without assuming $\nabla\cdot \bfv=0$.  To see this, we use  \eqref{eq:Peul} to obtain
\bq
\mathcal{P}_{\rho}\frac{\delta H}{\delta\mathbf{v}}= \rho \mathbf{v} 
-\nabla\Delta_{\rho}^{-1}\nabla\cdot \mathbf{v}  
\qquad\mathrm{and}\qquad
 \mathcal{P}_{\rho}\frac{\delta P_i}{\delta \varv_j}= \rho \, \de_{ij}  \,, 
\label{dePH}
\eq
which when inserted into  \eqref{eq:dbkt-4} gives
\begin{eqnarray}
\left\{ P_i,H\right\} _{*} & = & -\int\! d^{3}x\,\bigg[\frac{\rho}{2}  \frac{\p |\bfv|^2}{\p x_i} 
+ \varv_i   \nabla \cdot 
\left( 
\rho \mathbf{v}  -  \nabla\Delta_{\rho}^{-1}\nabla\cdot \mathbf{v}
 \right) 
\nonumber \\
  &  & +\left[ 
  (\nabla\times\mathbf{v})\times 
   \left(  \rho \mathbf{v} -  \nabla\Delta_{\rho}^{-1}\nabla\cdot \mathbf{v} \right)
 \right]_i
 \nonumber \\
 &  & + (\nabla\cdot\mathbf{v})  
  \left[ \left( \rho \mathbf{v}  -  \nabla\Delta_{\rho}^{-1}\nabla\cdot \mathbf{v} \right)_i
 -\rho   \varv_i\right]
 \bigg]=0
 \,,
 \label{PHbkt}
\end{eqnarray}
as expected.  The result of \eqref{PHbkt} follows upon using standard vector identities, integration by parts, and the self-adjointness of $\Delta_{\rho}^{-1}$.

The associated generator of  space translations that satisfies  the constraints is  given by  the operator $\{\bfP, \ \cdot\ \}_*$, which can be shown directly.  And, it follows that 
\bq
\{P_i,P_j\}_*= 0\,, \qquad\ \forall\  i,j=1,2,3\,.
 \label{PPbkt}
\eq
Because the momentum contains no $s$ dependence the the second line of \eqref{eq:dbkt-4} vanishes and using  $\mathcal{P}_{\rho}{\delta P_i}/{\delta \varv_j}= \rho \, \de_{ij}$ of  \eqref{dePH} it is clear the last line involving $\nabla\cdot \bfv$  of \eqref{eq:dbkt-4} also vanishes.  The result of \eqref{PPbkt} is obtained because the first and third lines cancel.

Next, consider  the   angular momentum 
\bq
\bfL=\int\! d^3x \, \rho\, \bfx\times\bfv\,.
\label{angMom}
\eq
We will show  
\bq
\left\{ L_i,H\right\} _{*} =0\,.
\label{LH}
\eq 
Using $\mathcal{P}_{\rho} {\delta L_i}/{\delta \bfv}={\delta L_i}/{\delta \bfv}$,  {which follows from \eqref{eq:Peul}  with  $\p(\ep_{ik\ell} x_\ell)/\p x^\ell=0$, }
the fact  that $\{L_i, H\}=0$ for the compressible fluid, and $ \mathcal{P}_{\rho}= I- \mathcal{P}_{\rho\perp}$, we obtain 
\begin{eqnarray}
\left\{ L_i, H\right\} _{*} & = & \int\! d^{3}x\,\Bigg[
- \nabla\frac{\delta L_i}{\delta\rho}\cdot\mathcal{P}_{\rho\perp}(\rho\bfv)
 \nonumber \\
 &  & +\left( 
 \frac{\delta L_i}{\delta\mathbf{v}}\times  \frac{\nabla\times\mathbf{v}}{\rho}  
+ \frac{\nabla\cdot \bfv}{\rho}\,  \frac{\delta L_i}{\delta\mathbf{v}} 
 \right) 
 \cdot   \mathcal{P}_{\rho\perp} (\rho\bfv) 
 \Bigg]\,.
 \label{LH1}
\end{eqnarray}
Next, recognizing that $\mathcal{P}_{\rho\perp}(\rho\bfv)= \nabla \De_\rho^{-1} \nabla\cdot\bfv$ and integrating by parts, we obtain
\bq
\left\{ L_i, H\right\} _{*}  =  \int\!d^{3}x\, \De_\rho^{-1}( \nabla\cdot\bfv)\,  \Bigg[
\nabla^2\frac{\delta L_i}{\delta\rho} 
 - \nabla\cdot \left( 
 \frac{\delta L_i}{\delta\mathbf{v}}\times  \frac{\nabla\times\mathbf{v}}{\rho}  
+ \frac{\delta L_i}{\delta\mathbf{v}} \,  \frac{\nabla\cdot \bfv}{\rho}
 \right) 
 \Bigg]\,.
 \label{LH2}
\eq
Then upon inserting
\bq
\frac{\delta L_i}{\delta \rho}= \ep_{ijk}x_j\varv_k
\qquad\mathrm{and}\qquad  \frac{\delta L_i}{\delta \varv_j}= \rho \, x_k\ep_{ikj}  \,,
\label{deLH}
\eq
 and using standard vector analysis we obtain \eqref{LH}.

Because  $\mathcal{P}_{\rho} {\delta L_i}/{\delta \bfv}={\delta L_i}/{\delta \bfv}$, the first and third lines of \eqref{eq:dbkt-4} produce 
\bq
\{L_i,L_j\}_*= \ep_{ijk} L_k \,,
\eq
just as they do for the compressible fluid (and MHD), while the fourth line manifestly vanishes.  Similarly, it follows that  that 
$\{\bfL,\cdot\, \}_*$ is the generator for rotations. 

To obtain the full algebra of invariants we need $\{L_i,P_j\}_*$.  However because $\mathcal{P}_{\rho} {\delta P_i}/{\delta \bfv}={\delta P_i}/{\delta \bfv}$ and $\mathcal{P}_{\rho} {\delta L_i}/{\delta \bfv}={\delta L_i}/{\delta \bfv}$, it follows as for the compressible fluid that  $\{L_i, P_j\}_*= \ep_{ijk} P_k$.

Finally, consider the following measure of the position of the center of mass, the generator of Galilean boosts,  
\bq
\bfG= \int\! d^3x\, \rho\, (\bfx - \bfv t)\,.
\eq
Calculations akin to those above reveal
\bq
\{G_i, G_j\}_* = 0\,,\quad \{G_i, P_j\}_* = 0\,, \quad \{G_i, H\}_* = P_i\,, \quad \{L_i, G_j\}_* =\ep_{ijk} G_k\,.
\eq

Thus the bracket \eqref{eqn:PBD} with the set of  ten invariants $\{H, \bfP,\bfL, \bfG\}$ is at once a closed subalgebra of Poisson bracket realization on all functionals  and produces an operator realization of the Galilean group \citep[see e.g.][] {sudarshan} that  is homomorphic to the operator algebra of $\{L_i, \ \cdot\ \}_*$, $\{P_i , \ \cdot\ \}_*$, etc.\ with operator commutation relations.  This remains true for MHD with the only change being the addition of $H_B$ to the Hamiltonian.

Thus, the Galilean symmetry properties of the ideal fluid and MHD are not affected by the compressibility constraint.   However, based on past experience with advected quantities,   we do expect a new Casimir invariant of the form
\bq
\hat C[\rho,s]=\int\! d^3x   \,\hat\calc(\rho,s)\,. 
\label{hatCas}
\eq
To see that   $\{\hat C, F\}_*=0$ for any functional $F$,  where $\hat\calc(\rho,s)$ is an arbitrary function of its arguments, we calculate  
\bq
\{F, \hat C\}_*=  -\int\!d^{3}x\,\frac1{\rho}
\left[
\rho\, \nabla \frac{\p \hat \calc}{\p \rho}  -   \frac{\p \hat \calc}{\p s}\, \nabla s
\right]
\cdot \calp_\rho \frac{\delta F}{\delta\mathbf{v}}\,,
\label{hatCas2}
\eq
and since $\nabla \times (\rho\, \nabla  {\p \hat\calc}/{\p \rho}  -    {\p \hat\calc}/{\p s}\, \nabla s)=0$ we write it as $\nabla p$,  giving for \eqref{hatCas2}  
\bq
\{F, \hat C\}_*=  -\int\! d^{3}x\,\frac1{\rho}
\nabla p
\cdot \calp_\rho \frac{\delta F}{\delta\mathbf{v}}\,.
\label{hatCas3}
\eq
Thus,  integration by parts and   use of  \eqref{divP} imply $\{F, \hat C\}_*=0$ for all functionals $F$.  Note, without loss of generality we can write $\hat \calc(\rho,s)= \rho U(\rho, s)$,  in which case $p=\rho^2 \p U/ \p \rho$.  Thus,  it is immaterial whether or not one retains the internal energy term $ \int\! d^3x\,  \rho U(\rho,s)$  in  the Hamiltonian. 

Now, \eqref{hatCas} is not the most general Casimir.  Because both $\rho$ and $\nabla\cdot\bfv$ are Lagrangian pointwise Dirac constraints,  we expect the following to be an Eulerian Casimir
\bq
\hat C[\rho,s,\nabla\cdot\bfv]=\int\! d^3x   \, \calc(\rho,s,\nabla\cdot\bfv)\,, 
\label{newCas}
\eq
where $\calc$ is an arbitrary function of its arguments.   To see that   $\{ C, F\}_*=0$ for any functional $F$, we first observe that
\bq
\frac{\de C}{\de \bfv}= -\nabla \frac{\p \calc}{\p \nabla\cdot \bfv}
\eq
and, as is evident from \eqref{eq:Peul},  that $\nabla\cdot (\calp_\rho\nabla\Phi)=0$ for all $\Phi$; hence, all the ${\de C}/{\de \bfv}$ terms vanish except the first term of the last line of \eqref{eq:dbkt-4}.  This term combines with the others to cancel, just as for the calculation of  $\hat\calc$.

For constant density, entropy, and magnetic field, the bracket of \eqref{eqn:PBD} reduces to
\bq
 \{F,G\}_*= -\int \!d^3x\,   \frac{\nabla\times {\bf v}}{\rho}\cdot \left( \P\frac{\de F}{\de \bf v}\times \P\frac{\de G}{\de \bf v}\right)\,, 
   \label{eqn:RPBD}
\eq
whence it is easily seen that the helicity
\bq
C_{\varv\cdot\nabla\times \varv}=\int\! d^3x\, \bfv\cdot\nabla\times \bfv
\eq 
is a Casimir invariant because   $\P\, (\nabla\times\bfv)= \nabla\times\bfv$. 
This Casimir is lost when entropy and density are allowed to be advected, for it is no longer a Casimir invariant of \eqref{eq:dbkt-4}.

Now, let us  consider invariants in the Lagrangian description. Without the incompressibility constraints,  the Hamiltonian has a standard kinetic energy term and the internal energy depends on $\p q/\p a$, an infinitesimal version of the two-body interaction,  if follows that just like the $N$-body problem the system has Galilean symmetry, and because the Poisson bracket in the Lagrangian description \eqref{cbkt} is canonical there are no Casimir invariants.  With the incompressibility constraint, the generators of the algebra now respect the constraints,  with Dirac constraints being  Casimirs and the algebra of constraints now having  a nontrivial center.  Because the Casimirs are pointwise invariants, we expect the situation to be like that for the Maxwell Vlasov equation \cite{pjm82}, where the following is a Casimir
\bq
C_{\nabla\cdot B}[\bfB]=\int \!d^3x\,  \calc(\nabla\cdot \bfB, \bfx)\,,
\label{divB}
\eq
with $\calc$ being an arbitrary function of its arguments.  {Because both nabla $\nabla \cdot \bfB$ and $\calj$ are pointwise constraints,  analogous to \eqref{divB} we expect   the following Casimir:}
\bq
\hat C[\calj]=\int \!d^3a\,  \hat\calc(\calj, \bfa)\,.
\eq
Indeed, only the first term of  \ \eqref{eq:dbkt-2} contributes when we calculate $\{\hat C, G\}_*$ and this term vanishes by \eqref{Ldiv} because
\bq
\frac{\de \hat C}{\de q^i}=-\frac{\p  }{\p a^\ell} \left( A^\ell_i\frac{\p \hat\calc}{\p \calj}\right)\,,
\eq
 {which follows upon making use of \eqref{dedet}.} Similarly, it can be shown that the full Casimir is
\bq
\hat C[D^1,D^2]=\int \!d^3a\,  \hat\calc(D^1,D^2,\bfa)\,, 
\eq
a Lagrangian  Casimir consistent with  \eqref{newCas}. 

For MHD, the  magnetic helicity, 
\bq
C_{A\cdot B}= \int \!d^3 x \,  \bfA\cdot \bfB\, , 
\eq
where $\bfB=\nabla\times \bfA$ is easily seen to be preserved and a Casimir up to the  usual issues regarding gauge conditions and boundary terms \citep[see][] {finn}.
We know that the  cross helicity
\bq
C_{\varv\cdot B}= \int\!d^3 x \, \bfv\cdot \bfB\, ,
\eq
is a Casimir of the compressible barotropic MHD equations, and it is easy to verify that it is also a Casimir of  \eqref{eqn:PBD} added to \eqref{eqn:BPBD}, that is for uniform density.  However, it is not a Casimir for the case with advected density, i.e.,  for the bracket of \eqref{eq:dbkt-4} added to \eqref{PvBbkt}.

\section{Conclusions}
\label{sec:conclusion}

In this paper we have substantially investigated constraints, particularly incompressibility for the ideal fluid and MHD,  for  the  three dichotomies described  in  Section \ref{ssec:bgnd}:  the Lagrangian vs.\ Eulerian fluid descriptions, Lagrange multiplier vs.\ Dirac constraint methods, and Lagrangian vs.\ Hamiltonian formalisms.  An in depth description of the interplay between the various fluid and MHD descriptions was given,  with an emphasis on Dirac's constraint method.  Although we mainly considered geodesic flow for simplicity, the Dirac's Poisson bracket method can be used to find other forces of constraint in a variety of fluid and plasma contexts.  

Based on our results, many avenues for future research are presented.  We mention a few.  Since the Hamiltonian structure of extended and relativistic MHD are now at hand \citep{pjmKLWW14,AKY15,pjmDP15,pjmLM16,pjmDL16,pjmKT20} calculations analogous to those presented here can be done for a variety of magnetofluid models.  {Another valuable class of models that could be studied, ones that  are known to have  Lagrangian and Hamiltonian structure, are those with various finite-Larmor-radius effects \citep[e.g.][]{pjmTWG08,izacard11,tassi14,tassi19} }

 Another avenue for future research would be to  address stability with constraints. In a previous series of papers \citep{pjmAP10,pjmAP12,pjmAP13,pjmAPE15,pjmAP16} we have investigated Hamiltonian  based stability, generalizations of the MHD energy principle or the ideal fluid Rayleigh criterion,  within the Lagrangian, energy-Casimir, and dynamically accessible frameworks.   Because Dirac's method adds Casimirs, the Dirac constraints, one gets a richer set of equilibria from the energy-Casimir variational principle and these can be tested for Lyapunov stability.  Similarly, the method of dynamical accessibility \citep[see][]{pjm98} based on constrained variations induced by the Poisson operator will enlarge the set of stable equilibria.   
 
  {Recently there has been consider research in the development of structure preserving computational algorithms.  \citep[See, e.g.,][for review.]{pjm17}  These are algorithms that preserve various geometric, Hamiltonian, variational,  and other structure of fluid, kinetic, and other physical models.  In the plasma community, in  particular,   we mention  \citet{evstati13,hong16,pjmXQLYZH16,pjmKKS17}, but there is a large body of  additional work  by these and other authors.  Given how the finite-dimensional material of Section \ref{sec:constraints} so strongly parallels the infinite-dimensional material of  Section \ref{sec:dirac},  notably the structure of geodesic flow,  a natural avenue for future research would be to develop numerical algorithms that preserve this structure.}

 Lastly, we mention that there is considerable geometric structure behind our calculations  {that could be further developed.  Our results can be restated }in geometric/Lie group language  \citep[see e.g.][]{bloch}.  Also, Arnold's  program for obtaining the Riemann curvature for geodesic flow on the group of volume preserving diffeomorphisms can be explored beginning from our results of Section \ref{sec:dirac}.  We did not feel this special issue would be the appropriate place to explore these ideas.

 
\section*{Acknowledgment}

\noindent PJM was supported by U.S. Dept.\ of Energy  under contract \#DE-FG02-04ER-54742.  He would also like to acknowledge support from the Humboldt Foundation and the hospitality of the Numerical Plasma Physics Division of the IPP, Max Planck, Garching.  FP would like to acknowledge  the hospitality of the Institute for Fusion Studies of the University of Texas at Austin.

\bibliographystyle{jpp}

\end{document}